\documentclass[APA,LATO1COL]{WileyNJD-v2}
\usepackage{moreverb}
\usepackage{graphicx,subfigure,amsthm,amsmath,latexsym,amssymb,times, caption}
\usepackage{float,epsfig,multirow,rotating,times,verbatim,wrapfig,color}
\articletype{Article Type}%

\usepackage[algo2e, ruled, vlined]{algorithm2e}
\received{26 April 2016}
\revised{6 June 2016}
\accepted{6 June 2016}

\RequirePackage{natbib}
\evensidemargin \oddsidemargin

\newcommand{\R}{I\!\!R}

\newcommand{\btheta}{\boldsymbol{\theta}}

\newcommand{\bSigma}{\boldsymbol{\Sigma}}

\newcommand{\bpsi}{\boldsymbol{\psi}}

\newcommand{\bmu}{\boldsymbol{\mu}}

\newcommand{\bTheta}{\boldsymbol{\Theta}}
\newcommand{\bvartheta}{\boldsymbol{\vartheta}}

\newcommand{\bh}{\boldsymbol{h}}

\newcommand{\bt}{\boldsymbol{t}}

\newcommand{\by}{\boldsymbol{y}}
\newcommand{\bY}{\boldsymbol{Y}}
\newcommand{\bZ}{\boldsymbol{Z}}

\newcommand{\mN}{\mathcal N}

\newcommand{\apprasym}{
 \mathrel{\ooalign{$\sim$\cr\kern+1.25pt\large $\colon$}}}

\begin{document}

\title{Learning trends of COVID-19 using semi-supervised clustering}

\author[1]{Semhar Michael*}
\author[2]{Xuwen Zhu}
\author[2]{Volodymyr Melnykov}

\authormark{Michael, Zhu, and Melnykov}

\address[1]{\orgdiv{Mathematics and Statistics Department}, \orgname{South Dakota State University}, \orgaddress{\state{Brookings}, \country{USA}}}

\address[2]{\orgdiv{Department of Information Systems, Statistics, and Management Science}, \orgname{The University of Alabama}, \orgaddress{\state{Alabama}, \country{USA}}}

\corres{*Semhar Michael, \email{semhar.michael@sdstate.edu}}

\presentaddress{Mathematics and Statistics Department, South Dakota State University, Brookings, SD 57007, USA}

\abstract[Summary]{A finite mixture model is used to learn trends from the currently available data on coronavirus (COVID-19). Data on the number of confirmed COVID-19 related cases and deaths for European countries and the United States (US) are explored. A semi-supervised clustering approach with positive equivalence constraints is used to incorporate country and state information into the model. The analysis of trends in the rates of cases and deaths is carried out jointly using a mixture of multivariate Gaussian non-linear regression models with a mean trend specified using a generalized logistic function. The optimal number of clusters is chosen using the Bayesian information criterion. The resulting clusters provide insight into different mitigation strategies adopted by US states and European countries. The obtained results help identify the current relative standing of individual states and show a possible future if they continue with the chosen mitigation technique.}


\jnlcitation{\cname{%
\author{S. Michael}, 
\author{X. Zhu}, and 
\author{V. Melnykov}} (\cyear{2020}), 
\ctitle{Learning trends of COVID-19 using semi-supervised clustering}, \cjournal{}, \cvol{}.}

\maketitle

%\footnotetext{\textbf{Abbreviations:} ANA, anti-nuclear antibodies; APC, antigen-presenting cells; IRF, interferon regulatory factor}

\section{Introduction}
\label{sect.intro}
 
According to the World Health Organization (WHO) as of April 28, 2020, 213 countries have reported close to 3 million cases and more than 202 thousand deaths due to the novel coronavirus disease (COVID-19), a respiratory tract infection \citep{who20}. The virus was first reported in Wuhan city, Hubei province of China on December 31, 2019. WHO declared this an outbreak of international public health emergency on January 31 and a pandemic on March 11, 2020. As the spread of the virus widened, different countries and territories reacted to the pandemic in various ways. The response generally starts by attempting to contain the virus by quarantining infected individuals, then moving to mitigation when containment is not possible. Without any intervention, it is estimated that around 40 million deaths worldwide \citep{walkeretal20}.  Governments have used many mitigation strategies ranging from those that implemented strict shelter-in-place orders ({\it e.g.}, China, Italy) to those that are advising social distancing without other substantial restrictions ({\it e.g.}, Netherlands, Sweden), with some targeting herd immunity \citep{dogruetal16, gardneretal20, fergusonetal20}. Another important factor in the spread of this virus is the availability and use of diagnostic tests \citep{whoreport8520}. Countries had different timelines for the availability of diagnostic tests and implemented different criteria on when and who to test. Hence the numbers that are officially reported are a reflection of this difference \citep{whoreport8520}. As a result, there can be discrepancies in the officially reported numbers and the true numbers of cases and deaths due to this disease. For the reasons listed above and others, the trends reported by countries and states have varied significantly exhibiting heterogeneity even after some preprocessing and standardization of the data is done. One interesting and important question regarding this pandemic would be to see which regions have similar trends in the spread of the disease. More specifically, we are interested in investigating how the state-level trends within the United States (US) compare with some European countries that have encountered the virus before the US.

One of the flexible models capable of capturing heterogeneity with great interpretability is a finite mixture model \citep{mclachlanandpeel00}. Model-based clustering is a probabilistic modeling approach to finding groups of similar observations in heterogeneous data \citep{mclachlanandpeel00, banfieldandraftery93}. Model-based clustering can be accomplished through finite mixture models with an assumption that each mixture component represents a standalone cluster. Further, a mixture of nonlinear regression models deals with finding groups of similar observations and fitting a nonlinear regression model within each group, simultaneously \citep{wedelanddesarbo95, grunandleisch07}. %An applications of regression mixture model to clustering regions can be found in \cite{holzhauseretal18}.
If there is an additional restriction on the memberships of observations, semi-supervised mixture modeling can be employed. In this setting, there exists partial information about the equivalence of observations. In particular, some observations must be restricted to be in the same cluster and some others can be required to be in different clusters. Using the terminology in \cite{melnykovetal16}, these are called positive and negative equivalence constraints, respectively. In our setting, we know ahead of time that observations (the daily cumulative counts) coming from the same region must belong to the same cluster. Hence, semi-supervised clustering with positive equivalences will be developed. In the nonlinear regression setting, the mean of the model can take different forms. In this project, the mean will be modeled using a generalized logistic function \citep{richards59} which is a flexible growth model applicable in our framework.

The rest of the paper is organized as follows. Section~\ref{sect.meth} describes the data collection and cleaning followed by the development of the proposed methodology. Sections~\ref{sect.utility} shows the applicability of the generalized logistic function to modeling country and state-related trends. Section~\ref{sect.resl} discusses the results of the proposed semi-supervised clustering. Finally, a conclusion and limitations of the work are provided in Section~\ref{sect.disc}.

\section{Methodology}
\label{sect.meth}

\subsection{Data collection and preprocessing}
\label{sect.data}
In this analysis, the daily cumulative number of confirmed cases and deaths per region are used to find regions with similar trends. For country-related data, the R package {\it coronavirus}  \citep{krispin20} is used for the daily summary of COVID-19 data. This dataset provides daily confirmed cases and deaths by country. For the US, the New York Times GitHub page \citep{newyorktimes20} provides cumulative counts of confirmed cases and deaths at the state and county levels. In our analysis, we considered the fifty states in the US and the District of Columbia (we will refer to them as US states). In addition, twenty three European countries that have had the virus for a longer period and have implemented strategies to mitigate against COVID-19 similar to the US states are considered. The countries included in our analysis are the following: Austria, Belgium, Croatia, Denmark, Finland, France, Germany, Greece, Hungary, Ireland, Italy, Netherlands, Norway, Poland, Portugal, Romania, Serbia, Slovenia, Spain, Sweden, Switzerland, Turkey, United Kingdom. In addition to COVID-19 related data, the population of each country was gathered from The World Bank \citep{worldbank20} database. For the state-level data, the US Census Bureau database \citep{uscensus20} was used to gather the estimated population of each state for the year 2019. 

\begin{figure}[h]
\centering
%\begin{subfigure}{}
  \includegraphics[angle=0,totalheight=2.4in]{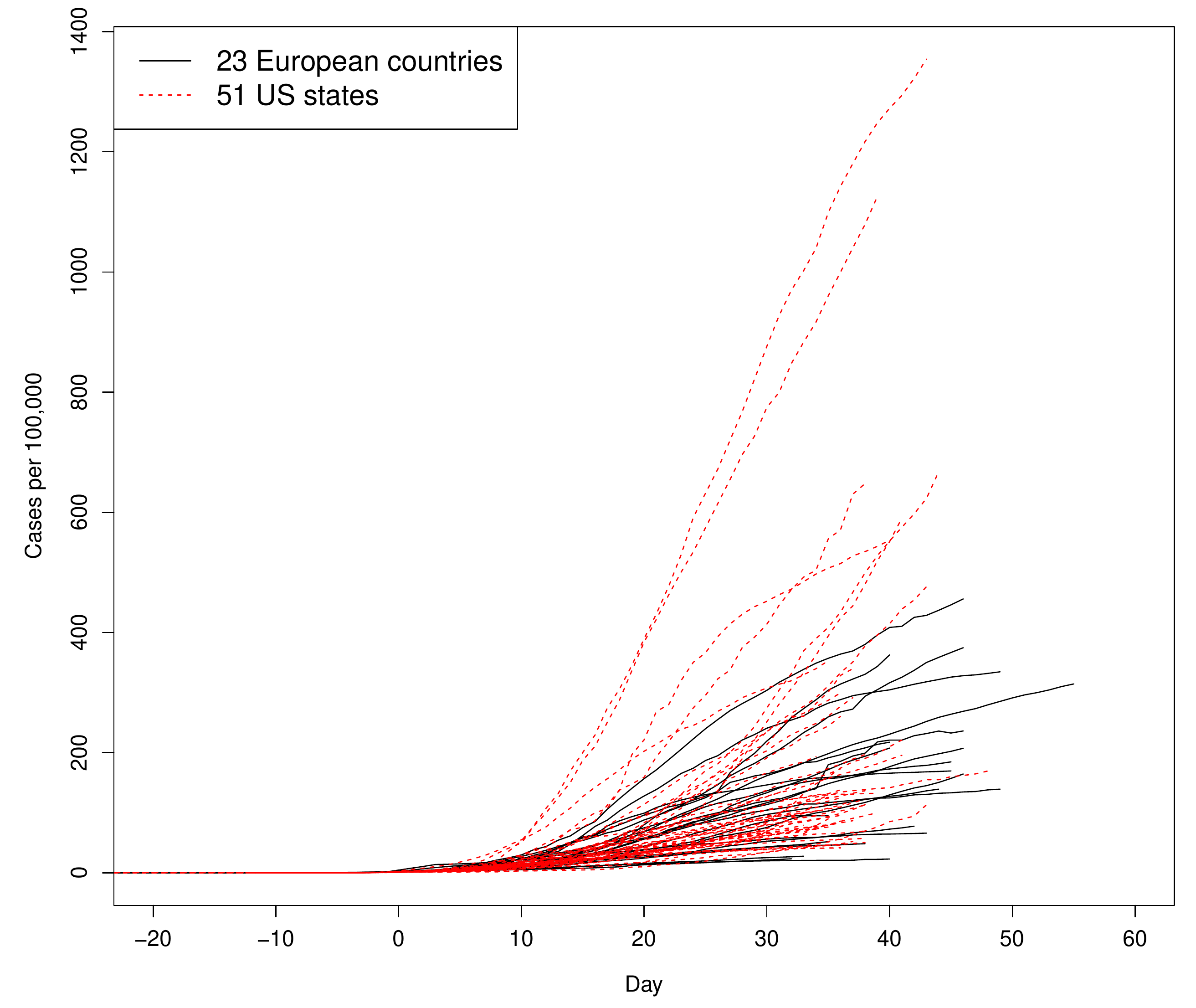}
  \includegraphics[angle=0,totalheight=2.4in]{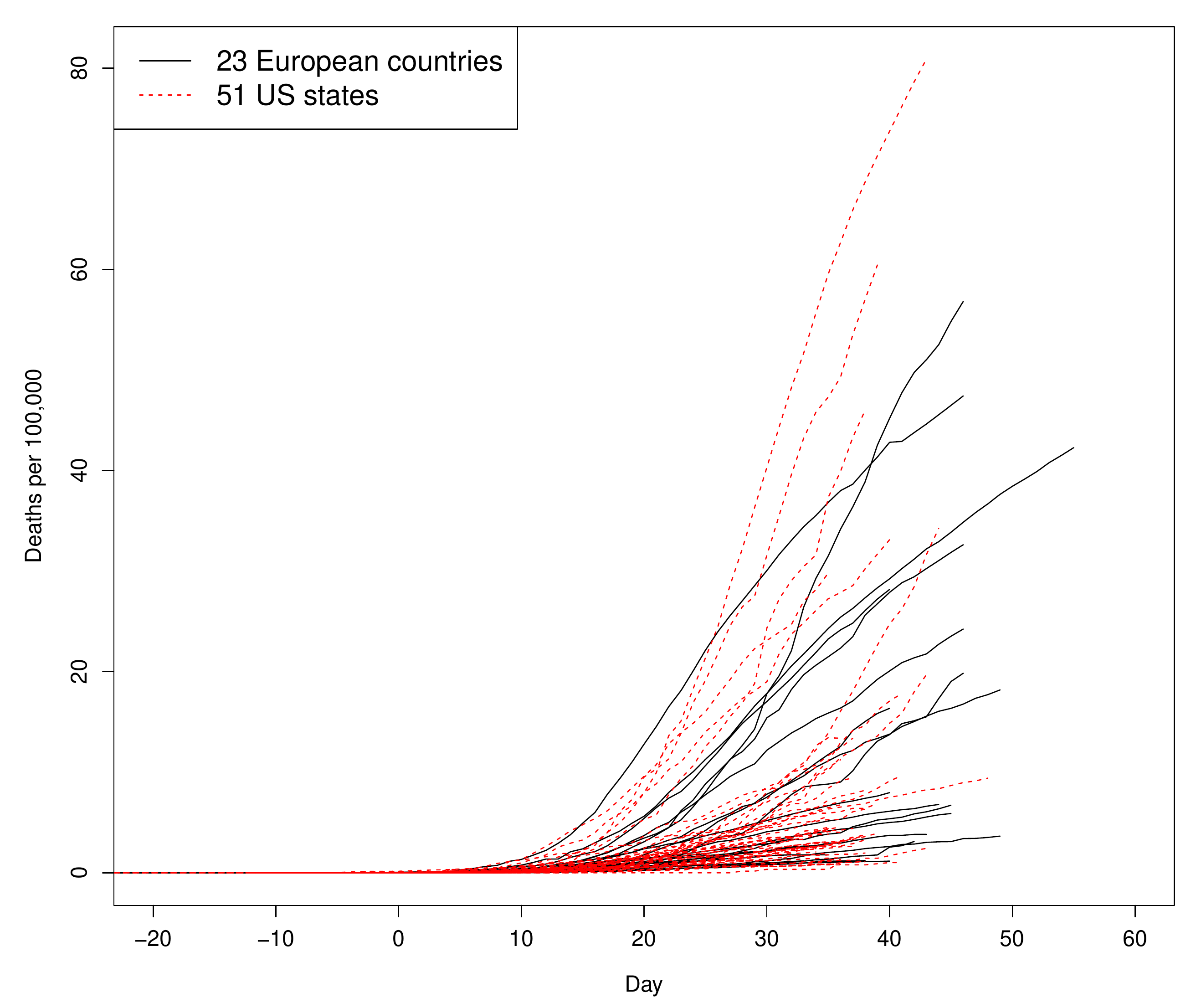}

\caption{Population adjusted cumulative counts aligned by the first date on which at least 1 case per 100,000 is reported (the time of onset).}
\label{fig.data}
\end{figure}

As a result, we have 74 regions included in our data (23 European countries and 51 US states). These regions differ in population size, therefore the data have been standardized by adjusting for population sizes as follows: $y_{bij}(t_{bi}) = x_{bij}(t_{bi}) / pop_i \times 100,000,$ for $b = 1, \ldots, 74$ regions, $i = 1, \ldots, n_b$ sequences of daily observations for the $b$th region, and $j = 1, 2$ cumulative confirmed cases and deaths at date $t$. In addition, each region reported its first case of the virus at different dates. However, the objective of this study is to identify similar trajectories in the development of the epidemic, regardless of the starting date. Therefore, we computed the first date that at least 1 case per 100,000 of the population was reported for each region ({\it i.e.,} $t_{b}^* = \min_i\{t_{bi}: y_{bi1}(t_{bi}) \geq 1\}$). Then, the regions were aligned according to this specific date, $t^*_b$. We will refer to this time point as {\it the time of onset} to represent an estimated time of the beginning of the epidemic for a given region. As a result, our analysis emphasizes on how the observed time series trends for each region compare aligned by this time of onset. Figure~\ref{fig.data} shows the sequence of a population adjusted cumulative cases and deaths for each region in our data. The figure is color- and line-type- coded to differentiate US states and European countries. In the figure, the value of zero in the x-axis indicates the time of onset. As we can see from the plot, most of the European countries are ahead of the US states. In our data, the state of West Virginia had the shortest time of 30 days and Italy had the longest time of 55 days after the time of onset. This data preprocessing allows for comparison of the regions since they are on the same scale in both rate of confirmed cases and deaths and the time variable. %Finally, to avoid negative values for time, the absolute value of the minimum value of time in the data was added to all sequences. 
Therefore, the modeling can be done to identify clusters in this collection of 74 regions.

\subsection{Mixture modeling and model-based clustering}
\label{sect.mbc}

Let $\{\by_i\}_{i=1}^n$ be a sample of $n$ independent observations identically distributed according to the probability density function (pdf) given by
\begin{equation}
  g(\by; \bTheta) = \sum_{k=1}^K \pi_k f_k(\by; \bpsi_k).
  \label{eq.fmm}
\end{equation}
Here, $\bTheta = (\pi_1, \ldots, \pi_K, \bpsi_1^\top, \ldots, \bpsi_K^\top)^\top$ is the full parameter vector, $\pi_1,\ldots,\pi_K$ represent mixing proportions or weights subject to constraints $0 < \pi_k \le 1$ and $\sum_{k=1}^K \pi_k = 1$, and $f_k(\by; \btheta_k)$ is the $k^{th}$ mixture component of known functional form with component-specific parameter vector $\btheta_k$. $K$ is the number of components, also known as the mixture order. The estimation of parameters is usually carried out by means of the expectation-maximization (EM) algorithm \citep{dempsteretal77}. The EM algorithm consists of two steps called E (expectation) and M (maximization) that are iterated until some pre-specified stopping criterion is met. The EM algorithm is known for its convenience in handling so-called missing information. In the mixture modeling framework, one assumes that the group labels $\{Z_i\}_{i=1}^n$ are missing. Then, the complete-data likelihood can be constructed based on the full data $\{\by_i, \bZ_i\}_{i=1}^n$. At the E step, the conditional expected value of the complete-data log likelihood function given observed data (traditionally denoted as the $Q$ function) is calculated. At the M step, the $Q$ function is optimized with respect to the parameter vector $\bTheta$. Upon convergence, the EM algorithm yields the maximum likelihood estimate $\hat\bTheta$. If the mixture order is unknown and has to be estimated, the most traditional approach is to employ one of information criterion, among which Bayesian information criterion (BIC) \citep{schwarz78} is the most popular in the mixture modeling context. BIC is calculated for a different number of mixture components and the model producing the lowest value is declared the winner.

A popular variant of finite mixtures is the mixture of regression models. In this setting, Gaussian pdf $\phi$ is usually employed for each component and the mixture is specified as
\begin{equation}
  g(y; \bt, \bTheta) = \sum_{k=1}^K \pi_k \phi(y; \mu_{k} (\bt) = h(\bt; \bvartheta_k), \sigma_k^2),
  \label{eq.fmrm}
\end{equation}
where $\sigma_k^2$ is the variance parameter and $\mu_k(t)$ is the mean of the $k^{th}$ component represented by the regression function $h(\bt; \bvartheta_k)$ with regression-specific parameters $\bvartheta_k$ and $\bt$ being the vector of explanatory variables. 

The most famous application of finite mixture modeling is model-based clustering that relies on the existence of the one-to-one relationship between mixture components and data groups. This relationship provides a highly intuitive interpretation of clustering results as data groups can be seen as samples from heterogeneous subpopulations constituting the superpopulation. The model-based clustering result is obtained based on the estimated posterior probabilities $\hat\tau_{ik}$ produced at the last E step of the EM algorithm. According to the Bayes decision rule, estimated membership labels are found as follows: $\hat Z_i = argmax_k \hat\tau_{ik}$.

\subsection{Generalized logistic function}
\label{sect.glf}

\begin{wrapfigure}{r}{0.4\textwidth}
\vspace{-0.5in}  
  \begin{center}
    \includegraphics[angle=0,totalheight=2.5in]{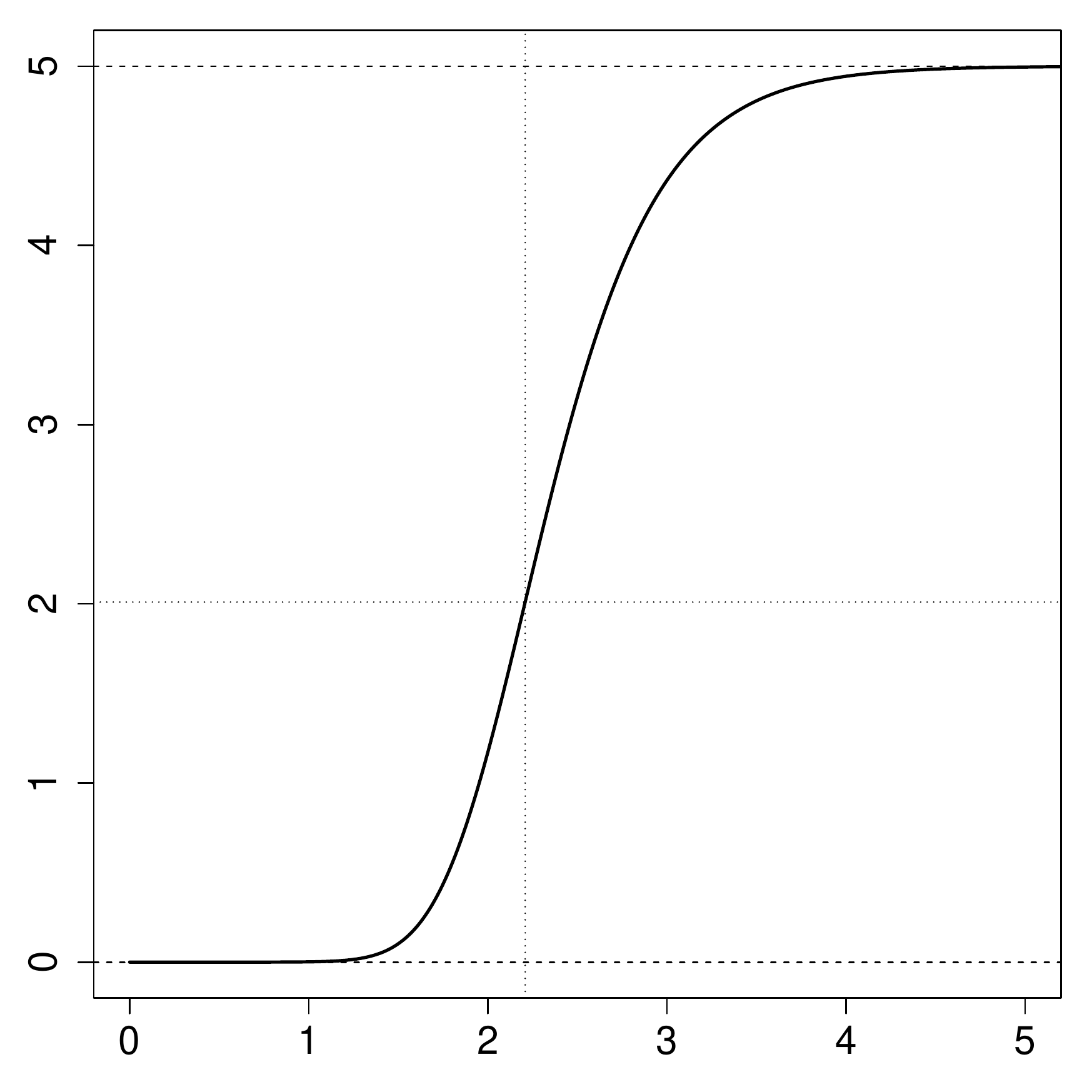}
  \end{center}
\vspace{-0.2in}
  \caption{Generalized logistic function.}
  \label{fig.glf}
  \vspace{-0.25in}
\end{wrapfigure}
Consider a nonlinear regression setting with $\{y_i\}_{i=1}^{n}$ denoting a sample of $n$ independent observations $y_i$ distributed according to normal distribution $\mN(h(t_i; \bvartheta), \sigma^2)$ with variance $\sigma^2$ and mean $h(t_i; \bvartheta)$. The functional form of the mean is chosen to be a four-parameter generalized logistic function defined as $h(t; \bvartheta) = a \big(1 + b e^{-ct}\big)^{-\gamma}$ for $t \in \R$ with $\bvartheta = (a, b, c, \gamma)^\top$ and $a, b, c, \gamma \in \R^+$. It is straightforward to notice that $0$ and $a$ are two horizontal asymptotes representing the lower and upper bounds of the function. Parameter $a$ is known as the carrying capacity, $b$ can be interpreted as a shift, and $c$ is the growth rate parameter. It can be shown that the inflection point is observed at $(t_0, y_0)$, where $t_0 = \log \{b\gamma\} / c$ and $y_0 = a (1+ \gamma^{-1})^{-\gamma}$. If $\gamma = 1$, the inflection point is located at $y_0 = a / 2$. Hence, parameter $\gamma$ is responsible for the shift of the inflection point toward the lower (if $\gamma > 1$) or upper (if $0 < \gamma < 1$) bound. The considered form of the generalized logistic function illustrated in Figure~\ref{fig.glf} presents a flexible tool for modeling the cumulative number of observed disease cases or deaths. In our framework, we model the total number of cases and overall number of deaths jointly. This leads to the multivariate nonlinear regression setting with bivariate response $\by = (y_{1}, y_{2})^\top$, where the first and second vector coordinates represent the cumulative numbers of reported disease cases and deaths, respectively. Then, the sequence of $n$ independent data points observed at times $\{t_i\}_{i=1}^n$ is given by $(2 \times n)$-dimensional matrix $\bY = \{\by_i\}_{i=1}^n$ with $\by_i \sim \mN_2(\bmu(t_i), \bSigma)$, where $\bmu(t_i) = \bh(t_i; \btheta)\equiv (h(t_{i}; \bvartheta_1), h(t_{i}; \bvartheta_2))^\top$ is the mean vector and $\bSigma$ is the covariance matrix. Vectors $\bvartheta_1$ and $\bvartheta_2$ represent variable specific parameters of the generalized logistic function and $\btheta = (\bvartheta_1^\top, \bvartheta_2^\top)^\top$.

\subsection{Semi-supervised clustering of regression mixture modles}
\label{sect.semi}

Let each of $B$ countries and states considered in this paper be represented by the observed response matrix $\bY_i$ of dimensions $2 \times n_b$, where $b = 1,\ldots,B$ and $n_b$ is the length of the $b^{th}$ bivariate sequence. In other words, we observe $\{\bY_b\}_{b=1}^B$ with $\bY_b = \{\by_{bi}\}_{i=1}^{n_b}$. The objective of this paper is to reveal common trends among heterogeneous data matrices $\{\bY_b\}_{b=1}^B$. However, the direct application of the mixture model
\begin{equation*}
  g(\by; t, \bTheta) = \sum_{k=1}^K \pi_k \phi_2(\by; \bh(t; \btheta_k), \bSigma_k),
\end{equation*}
where $t$ represents time and $\phi_2$ is the bivariate Gaussian pdf, is not possible in our context as data points observed at different times within the same country-specific sequence cannot be assigned to different components. In other words, there should be a condition that prevents assigning observations $\by_{bi}$ and $\by_{bi'}$ to different clusters.
To address this undesired feature of the model, one can employ so-called semi-supervised clustering with positive equivalence constraints. Such constraints assume that some observations in the data set are known to belong to the same group and hence must be treated jointly. In this framework, it is easier to work with blocks of observations tied by positive constraints. In our context this implies that each country or state represents a separate block with all elements required to belong to the same cluster. Then, the clustering procedure needs to be applied to blocks. 
%Let $\{\mB_b\}_{b=1}^B$ represent $B$ blocks, where each $\mB_b$ contains indices of objects, countries or states in out context, belonging to the $b^{th}$ block. By construction, $\cup_{b=1}^B \mB_b \equiv \{1,\ldots,n\}$ and $\mB_b \cap \mB_{b'} = \emptyset$ for any $b \ne b'$.
The EM algorithm corresponding to the semi-supervised clustering setting needs to take into consideration the fact that membership labels of observations belonging to the same block must be equal. For more details on positive and negative equivalence constraints, we refer the reader to the paper by \cite{melnykovetal16}.
It can be shown that the $Q$-function associated with our proposed model is given by
\begin{equation}
Q(\bTheta; \dot{\bTheta}, \{\bY_b\}_{b=1}^B) = \sum_{b=1}^B  \sum_{k=1}^K \ddot{\tau}_{bk} \sum_{i=1}^{n_b} \Big\{ \log \pi_k - \log 2\pi - \frac{1}{2} \log |\bSigma_k|
 -\frac{1}{2} \left(\by_{bi} - \bh(t_{bi}; \btheta_k)\right)^\top \bSigma_k^{-1} \left(\by_{bi} - \bh(t_{bi}; \btheta_k)\right)\Big\},
\label{eq3}
\end{equation}
where two dots on the top of posterior probabilities $\tau_{bk}$ represent the estimates at the current iteration of the EM algorithm, while one dot on the top of $\bTheta$ refers to the previous iteration.
It can be shown that the E step reduces to updating posterior probabilities by the expression
\begin{equation*}
 \ddot{\tau}_{bk} = \frac{{\dot{\pi}_k}^{n_b} \prod_{i=1}^{n_b} \phi_2 
\left(\by_{bi};\bh(t_{bi}; \dot\btheta_k), \dot\bSigma_k \right)
} {\sum_{k'=1}^{K} {\dot\pi_{k'}}^{n_b} \prod_{i=1}^{n_b} \phi_2 
\left(\by_{bi};\bh(t_{bi}; \dot\btheta_{k'}), \dot\bSigma_{k'} \right)}.
\end{equation*}
It is worth mentioning that posterior probabilities $\tau_{bk}$ refer to the chances of the entire block $b$ to belong to the $k^{th}$ component. The M step consists of the following expressions for updating parameters:
\begin{equation*}
\ddot\pi_k = \frac{\sum_{b=1}^B n_b \ddot\tau_{bk}}{\sum_{b=1}^B n_b} \quad \mbox{and} \quad \ddot\bSigma_k = \frac{\sum_{b=1}^B \ddot\tau_{bk} \sum_{i=1}^{n_b} \left(\by_{bi} - \bh(t_{bi}; \dot\btheta_k)\right)\left(\by_{bi} - \bh(t_{bi}; \dot\btheta_k)\right)^\top}{\sum_{b=1}^B n_b \ddot\tau_{bk}}.
\end{equation*}
Parameters associated with the generalized logistic functions can be updated by the numerical maximization of the function
\begin{equation*}
\tilde Q(\btheta_k) = \sum_{b=1}^B \ddot{\tau}_{bk} \sum_{i=1}^{n_b} \left(\by_{bi} - \bh(t_{bi}; \btheta_k)\right)^\top \ddot\bSigma_k^{-1} \left(\by_{bi} - \bh(t_{bi}; \btheta_k)\right)
\end{equation*}
with respect to $\btheta_k$ for $k = 1, \ldots, K$. This concludes the steps of the EM algorithm. 
 
\subsection{Computation aspects}
In our analysis, the stopping criterion is based on the relative difference in log likelihood values from two consecutive iterations. In addition, the initialization of the EM algorithm is crucial in obtaining a reasonable solution. In our analysis, we used multiple random starts approach for different values of $K$ and the best solution was chosen based on BIC. In particular, each candidate model $K>1$ was initialized with $M = \min\{K\times 20,100\}$ random starts. The EM algorithm is stopped in one of the two scenarios: 1) if the relative difference in log likelihood values from two consecutive iterations is less than a tolerance level of $10^{-6}$ or 2) if one or more spurious solutions are obtained. In the latter case, the solution is not retained. A Linux cluster "Roaring Thunder" at South Dakota State University is used to run the jobs for $M$ random starts at each $K$ in parallel. Looking forward, as the data on the disease are still being updated, the model can also be updated by initializing the EM-algorithm using the best model parameter estimates from the previous iteration of data.
\section{Results}
\label{sect.analysis}
%This section will address the question of how good the generalized logistic function fits the current set of data. Then the results of the semi-supervised clustering model will be presented. 

\subsection{The utility of the generalized logistic function}
\label{sect.utility}

\begin{figure}[h]
\scriptsize
\centerline{
  \mbox{
    \subfigure{\includegraphics[angle=0,totalheight=3.0in,width=3.2in]{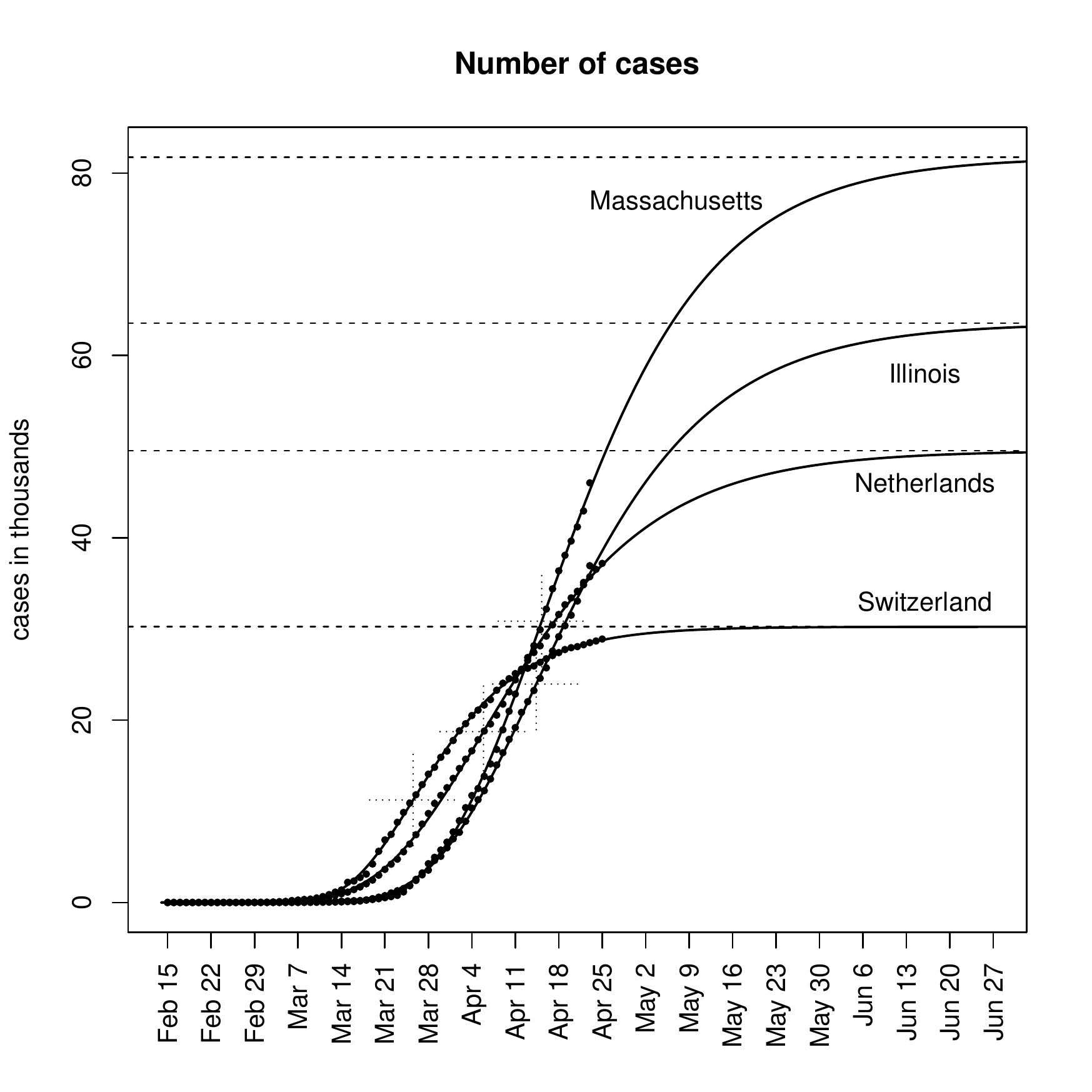}}
    \subfigure{\includegraphics[angle=0,totalheight=3.0in,width=3.2in]{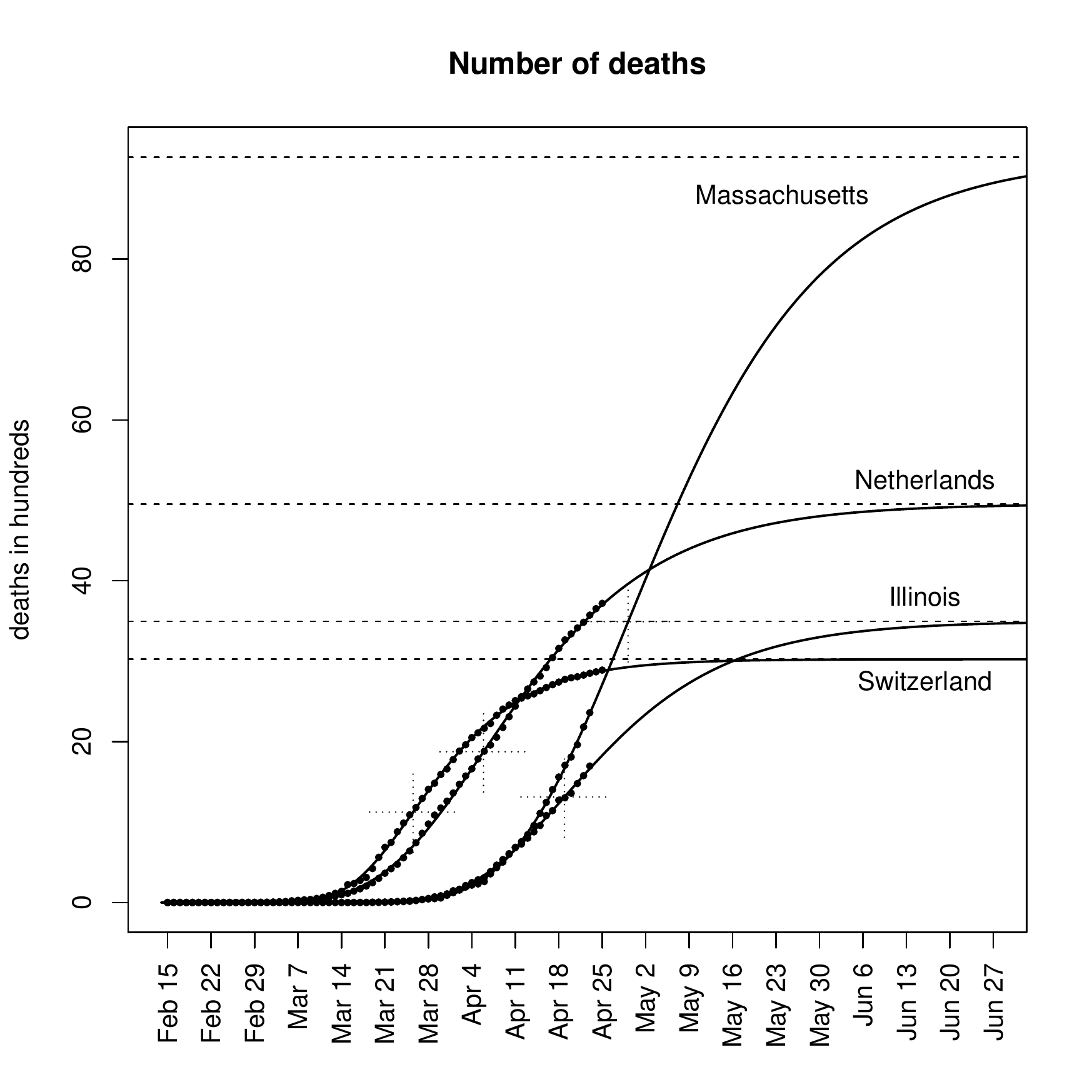}}     
  }}
\caption{Generalized logistic function fitted to four regions.}
\label{fig.countr}
\end{figure}

In this section, we discuss the utility of the generalized logistic function in the framework of the considered problem. Figure~\ref{fig.countr} shows plots constructed for cumulative numbers of disease cases (left plot) and deaths (right plot) for Switzerland, Netherlands, Illinois, and Massachusetts. The two countries are just two examples of European countries included in our study. Netherlands and Switzerland are not the first countries that have been hit with the virus and not the last in Europe either. The two US states have been chosen based on similar reasoning.

Each plot in Figure~\ref{fig.countr} presents inflection points reflected by crossed dotted lines and horizontal asymptotes (dashed lines) showing the expected total number of disease cases and deaths.
It can be noticed that the generalized logistic function fits the data very well in all cases. An important observation can be made about the onset time of the epidemic. Both European countries encountered a steady growth in the number of cases and deaths around March 14, while both states faced the disease extension roughly two weeks later in March. This motivates our paper as the analysis of European trends can be used for selecting an optimal public health policy for US states. Indeed trends observed for different countries and states are not identical. For example, Switzerland is considerably closer to the horizontal asymptote than the Netherlands and the two states.
The fitted curves can be particularly helpful for making decisions with regard to when quarantine measures can be relaxed. For example, Switzerland can is expected to overcome the outbreak by the middle of May while the Netherlands would roughly require an extra month to handle the disease. Expectedly, Illinois and Massachusetts would need even more time to approach the upper bound. It is important to mention that the fitted curves are based on current public health policies. If a country or state changes its response to the disease, for instance, by terminating the stay-at-home order, a different data trend will be observed and the generalized logistic function is likely to produce an unsatisfactory fit. For example, we observed such a trend for China after the quarantine was relaxed and new disease cases were primarily observed among visitors arriving from the outside of the country.

\newpage
\subsection{Clustering results}
\label{sect.resl}

\begin{wrapfigure}{r}{0.4\textwidth}
\vspace{-0.5in}  
  \begin{center}
    \includegraphics[angle=0,totalheight=2.0in]{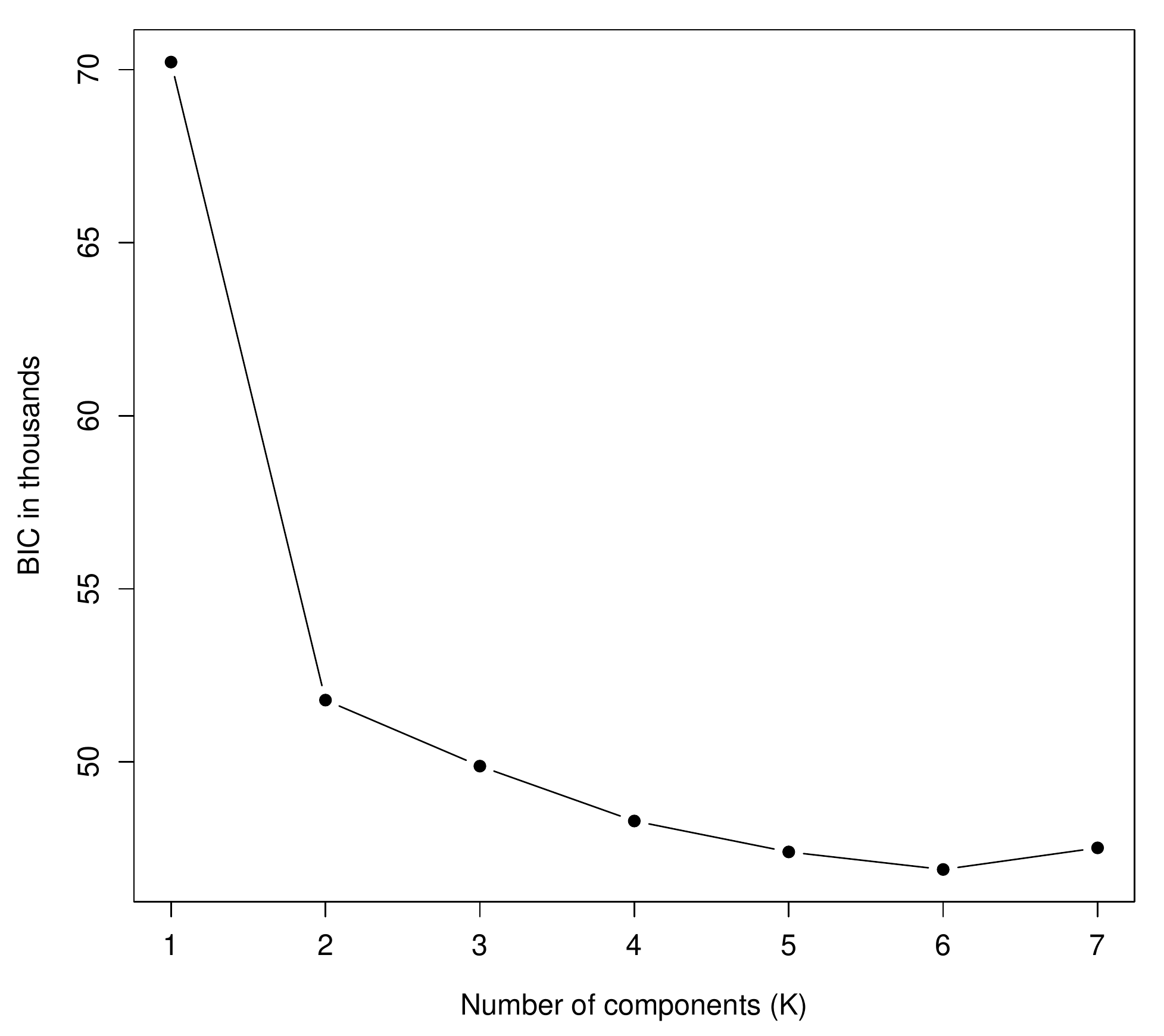}
  \end{center}
\vspace{-0.25in}
  \caption{Model selection using BIC.}
  \label{fig.bic}
  \vspace{-0.25in}
\end{wrapfigure}

The proposed clustering model was fitted to the data (74 total observations: 23 European countries and 51 US states) varying the number of clusters $K=1,\ldots, 7$. %The models with larger than seven clusters ($K>7$) did not result in any non-spurious solution.
BIC for $K=1,\ldots, 7$ can be seen in Figure~\ref{fig.bic}. The optimal model for the considered dataset based on BIC is the six-component mixture suggesting there are 6 distinct groups (Groups 1-6) in our dataset. The parameter estimates of the best model fit are reported in Table~\ref{table.pars}. The six groups are ordered in the table according to the expected upper bounds on the rate of cases in each group ($\hat{a}_{1k}$). 

\begin{figure}[ht]
\centering
  \includegraphics[angle=0,totalheight=2.4in]{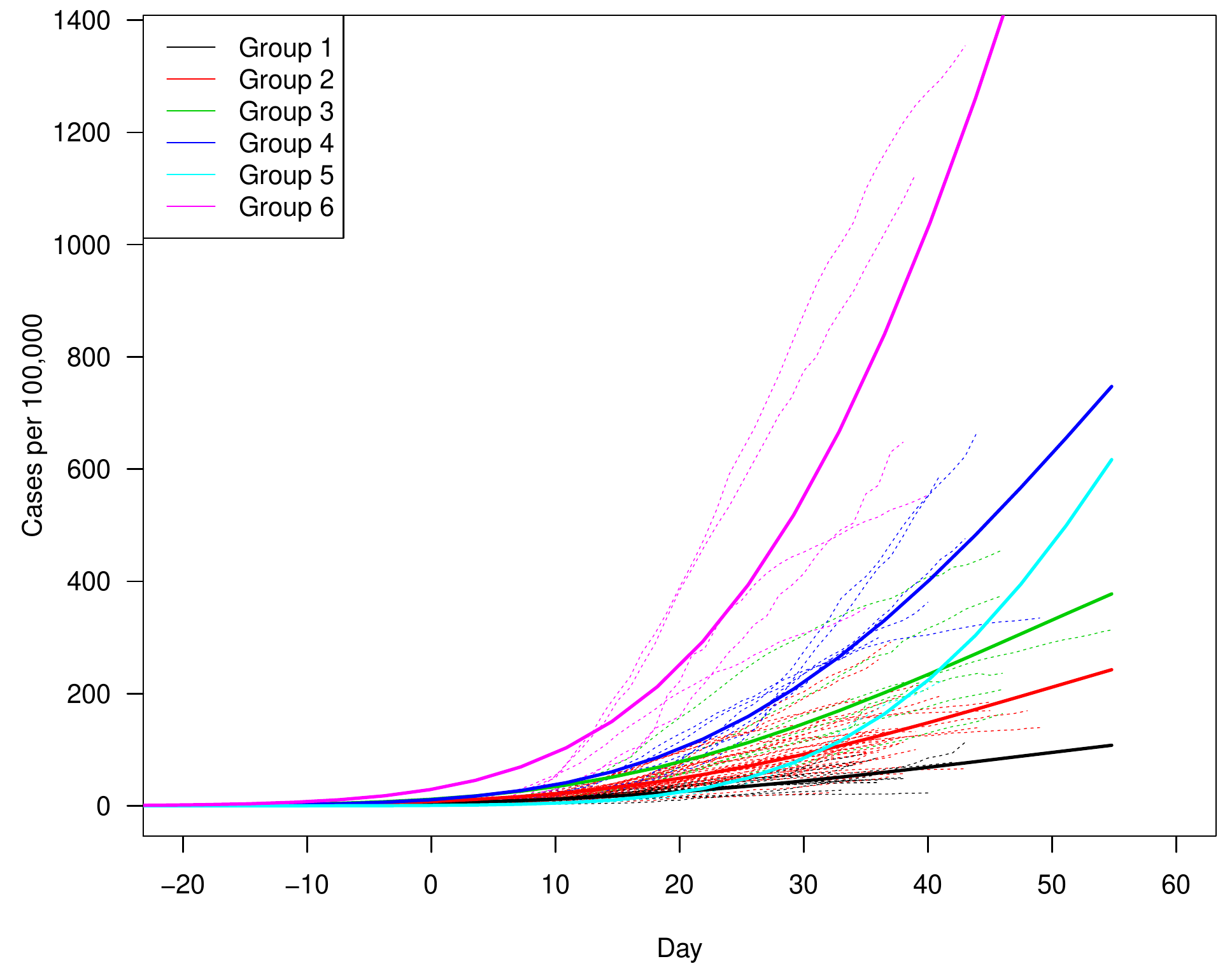}
  \includegraphics[angle=0,totalheight=2.4in]{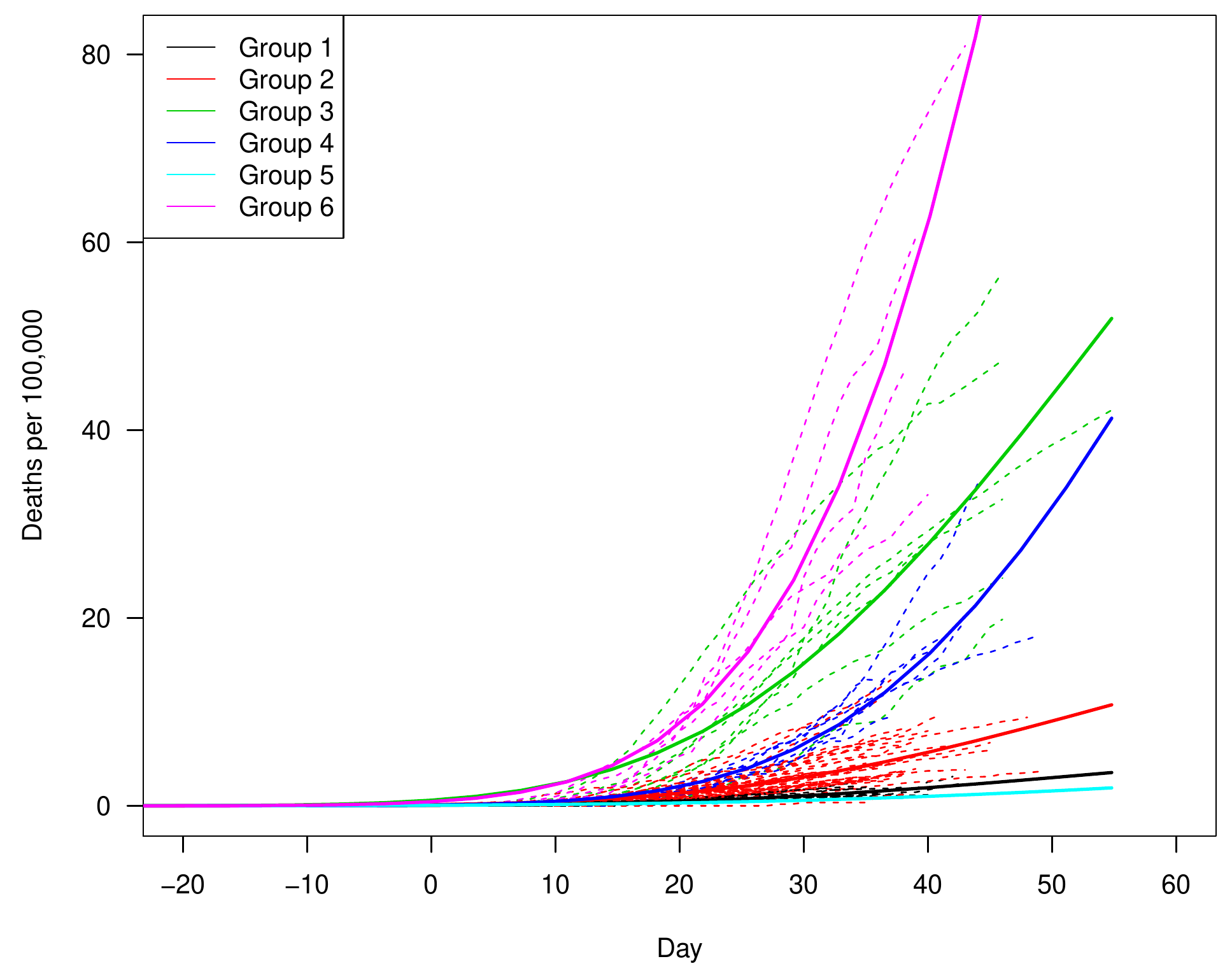}
  \caption{The six groups detected by the best model-based clustering solution.}
  \label{fig.clust}
  \vspace{-0.2in}
 \end{figure}

According to the best model, the upper bound of the number of cases ranges from 232 (in Group 1) to 7,497 (in Group 6) cases per 100,000. On the other hand, the number of deaths ranges from 9 (in Group 1) to 822 (in Group 6) per 100,000. Figure~\ref{fig.clust} shows a color-coded assignment to the six clusters and the corresponding cluster mean trend in each cluster. Individual cluster plots including the members of the cluster are given in Figure~\ref{fig.groups1}. The first column in this plot represents population-adjusted cases and the second column shows population-adjusted deaths. The legend of the plot lists the countries and US states that has been assigned to the group according to the highest posterior probability. The regions in the legend are listed in the descending order of the cumulative rates.

\begin{table}[ht]
\caption{\label{tab:}Parameter estimates}
\centering
\begin{tabular}{|c|c|c|c|c|c|c|c|c|c|c|c|c|}
\hline
$k$ & $\hat \pi_k$ & $\hat a_{1k}$ & $\hat a_{2k}$ & $\hat b_{1k}$ & $\hat b_{2k}$ & $\hat c_{1k}$ & $\hat c_{2k}$ & $\hat \gamma_{1k}$ & $\hat \gamma_{2k}$ & $\hat\sigma_{1k}$ & $\hat\sigma_{2k}$ & $\hat \rho{_k}$\\
\hline
\hline
1 & 0.189 & 232.003 & 9.206 & 3.809 & 4.983 & 12.197 & 12.930 & 8.185 & 9.747 & 8.814 & 0.203 & 0.613\\
\hline
2 & 0.541 & 554.105 & 29.341 & 3.640 & 4.567 & 12.263 & 12.866 & 9.396 & 10.925 & 23.421 & 1.048 & 0.756\\
\hline
3 & 0.095 & 817.711 & 140.018 & 3.069 & 3.998 & 12.475 & 12.714 & 11.007 & 11.746 & 44.120 & 5.054 & 0.925\\
\hline
4 & 0.095 & 2193.871 & 199.540 & 4.103 & 6.403 & 12.086 & 12.639 & 10.387 & 11.682 & 46.557 & 1.903 & 0.809\\
\hline
5 & 0.014 & 3217.663 & 10.916 & 6.510 & 1.887 & 12.858 & 8.364 & 12.819 & 12.215 & 8.087 & 0.094 & -0.036\\
\hline
6 & 0.068 & 7496.741 & 822.326 & 5.363 & 7.058 & 11.510 & 12.250 & 8.298 & 10.041 & 143.919 & 5.828 & 0.879\\
\hline
\end{tabular}
\label{table.pars}
\end{table}

\paragraph{Description of clusters}
\underline{Group 1 (black cluster)}: This group has 4 European countries (Finland, Croatia, Poland, Greece) and 11 US states (Nebraska, North Dakota, Arkansas, Texas, North Carolina, West Virginia, Oregon, Alaska, Hawaii, Montana) listed in the descending order of their current infection rate. The group has the lowest rate of confirmed cases and deaths as compared to the other groups.  In addition to that, the countries in this group have relatively shorter sequences with an average of 36 days since the time of onset (zero on the x-axis). Overall, the group appears to include Eastern European countries and US states that faced the disease at a later time. Besides, several of the US states within this group are characterized by low population density. The regions in this group should aim at maintaining this trend. %The groups expected upper bound is 2320 cases and 90 deaths per million.

\underline{Group 2 (red cluster)}: This group contains the largest number of regions with 10 European countries and 30 US states (see the list in Figure~\ref{fig.groups1}). The group has the second smallest rate of infection with a modest death rate. Relative to Group 1, this group has been in the epidemic for a slightly longer period with an average of 38 days since the time of onset. There are some interesting things we can note as we look at individual regions within the group. Illinois and Maryland are the top two in terms of the population-adjusted case and death rates in this group, while Utah has the lowest death rate even with a mid-level rate of cases. From the European countries in the group, Portugal has the highest rate of cases and deaths while Slovenia has low rate of cases with a mid-level rate of death. In general, this group represents trends observed in European countries handling the pandemic the best: Germany, Austria, Norway, Denmark, and some Eastern European countries that faced the pandemic later: Serbia, Hungary, Romania, and Turkey. It is of interest that the group includes relatively densely populated European countries such as Germany, Denmark, and Austria and sparsely populated states such as Wyoming, Utah, and Idaho. This could be due to the effective mitigating strategies adopted by the former.  The case of these European countries and US states such as California and Washington is to be mentioned as exemplary as they have large populations with high density and have had the pandemic for long periods but have ended up in a group with the second-best result.

\underline{Group 3 (green cluster)}: Group 3 is composed of 7 European countries and no US states. The countries in this group ordered by the rate of cases are Spain, Belgium, Italy, France, the United Kingdom (U.K.), Netherlands, and Sweden. This group contains countries in Europe that had encountered the disease earlier than most (with an average of 46 days after the time of onset). They are composed of countries that either are implementing limited restrictions (Netherlands, Sweden) or have implemented restrictions with some delay (Spain, Italy, and the United Kingdom). The group is characterized by high COVID-related mortality relative to the corresponding rate of detected diseases. Belgium and Spain are the top two countries in both rates of cases and death. It is interesting to notice that even though Spain has had a higher rate of confirmed cases for most of the duration, Belgium has shown much faster growth in terms of death rate after about 35 days since the time of onset. However, according to the official Belgian government site \citep{belgium20}, the country includes both confirmed and suspected deaths due to COVID-19 in the officially reported counts as opposed to other countries often reporting only confirmed cases. The US is currently considering reporting both confirmed and suspected cases separately for each US state \citep{newyorktimes20}. 

\underline{Group 4 (blue cluster)}: This group has two European countries: Ireland and Switzerland and five US states namely Massachusetts, Rhode Island, District of Columbia, Delaware, and Pennsylvania. Even though Group 4 has on average a higher rate of confirmed cases, the average death rate is lower than that of Group 3. An average of 41 days since the time of onset makes this group the second-longest since encountering the epidemic. In general, this group represents regions that are highly affected by COVID-19. Despite the very high disease rate, their mortality rate seems to be under control compared with that of countries in Group 3. 

\underline{Group 5 (cyan cluster)}: This group contains only one region - the state of South Dakota (SD). This state has the unique characteristic of having the second-highest rate of cases with an expected population-adjusted upper bound of 3217 and the lowest rate of deaths with a population-adjusted expected upper bound of 11. The reason for this specific trend might be that the high number of cases are concentrated in the Smithfield Foods Incorporation - a pork processing plant in Minnehaha county making the demography slightly younger than the at-risk group. Currently, the US age distribution shows that around 80\% of deaths are observed for individuals above the age of 65 \citep{cdcreport20}. On the other hand, 80\% of cases in South Dakota are from less than 60 years old individuals  \citep{sddoh20}. An additional reason for the low death rate might be the time lag from severe symptoms to death which is estimated to be close to two weeks \citep{wilsonetal20}. 
%The state currently has (439, 1) per 100,000 confirmed cases and deaths and is expected to reach (3218 and 11) per 100,000 cases and deaths, respectively.

\underline{Group 6 (magenta cluster)}: The last group had the largest number of cases and deaths. This group contains five US states namely New York, New Jersey, Connecticut, Louisiana, and Michigan arranged in the descending order of infection rates. This group is the most severely affected with no European country encountering the disease to such an extent. The group averaged 39 days since the time of onset with New York being the longest at 43 days. Both infection and mortality rates are higher than any other group with an expected upper bound of 7497 cases and 822 deaths per 100,000.

\begin{figure}[b!]
\centering
   \mbox{
  	\subfigure{\includegraphics[angle=0,totalheight=2.3in]{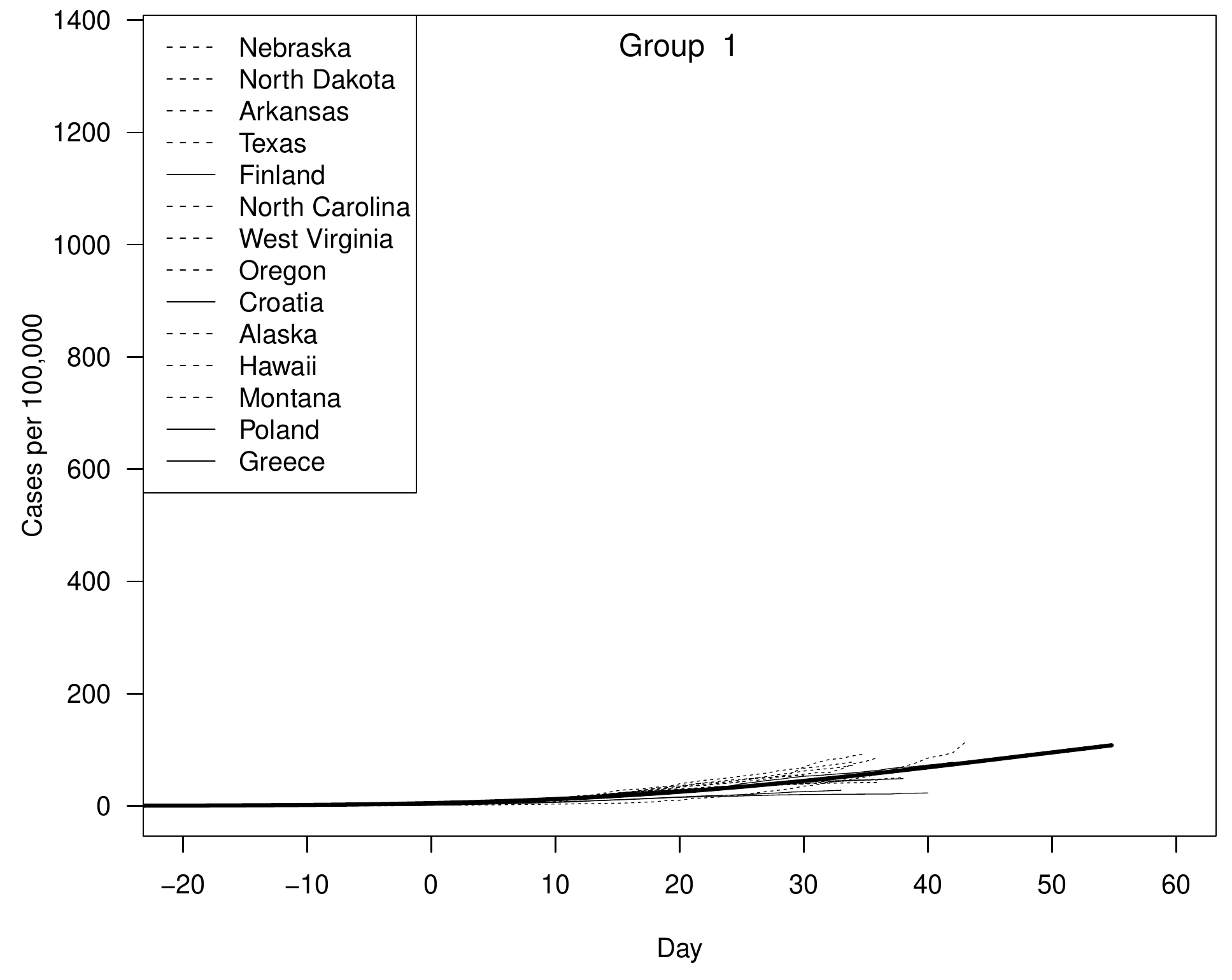}}
      	\subfigure{\includegraphics[angle=0,totalheight=2.3in]{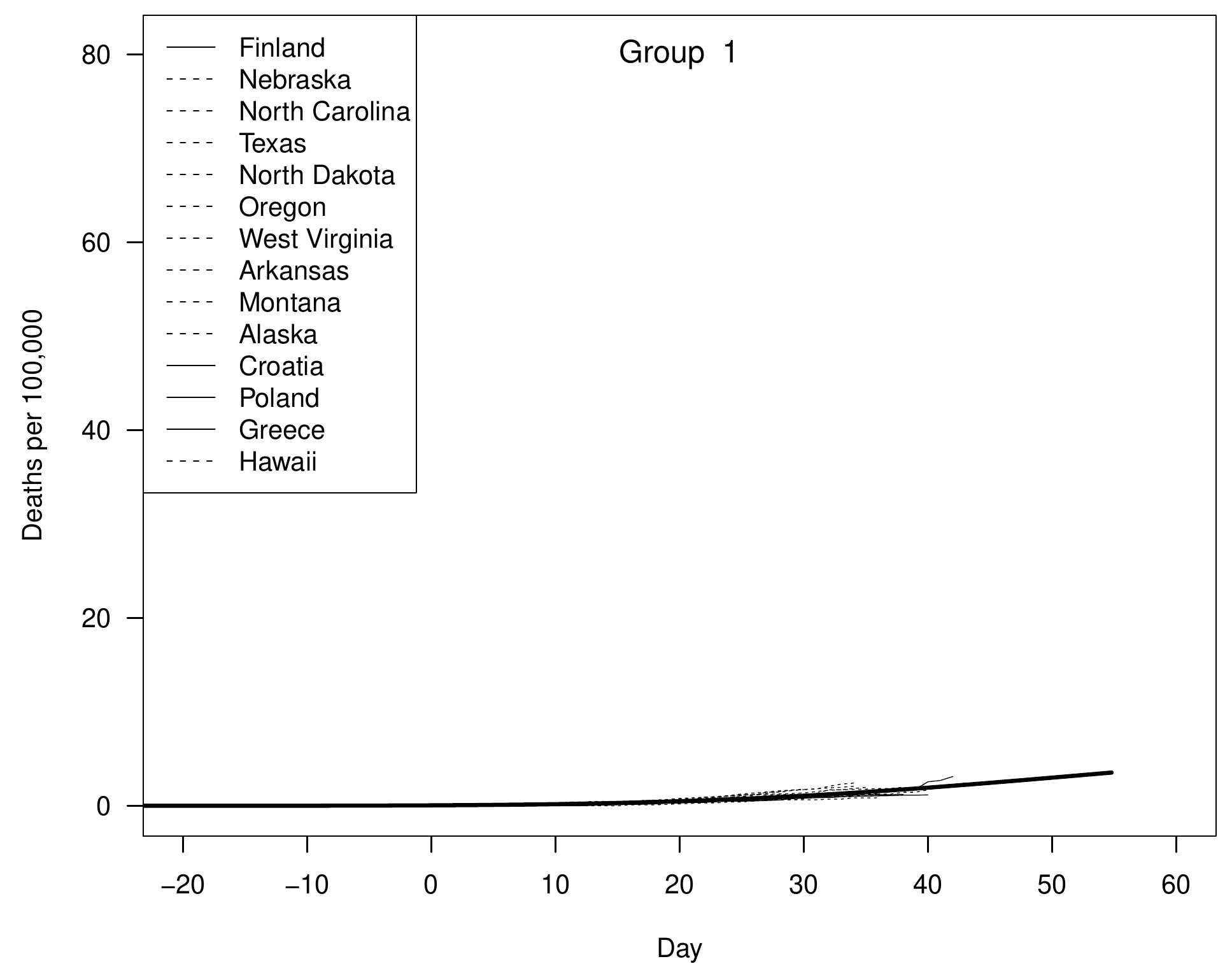}}
      }
  \mbox{
       	\subfigure{\includegraphics[angle=0,totalheight=2.3in]{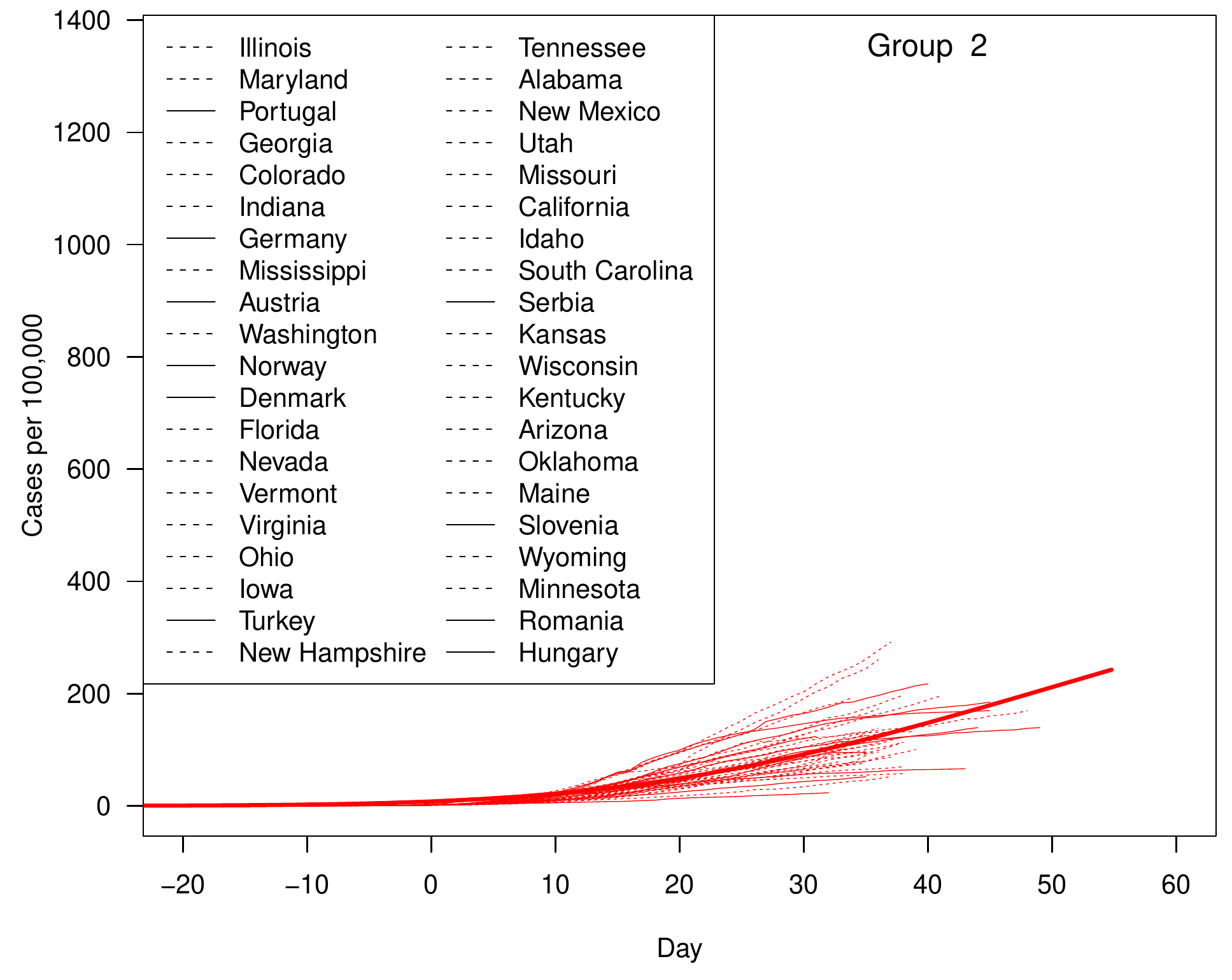}}
       	\subfigure{ \includegraphics[angle=0,totalheight=2.3in]{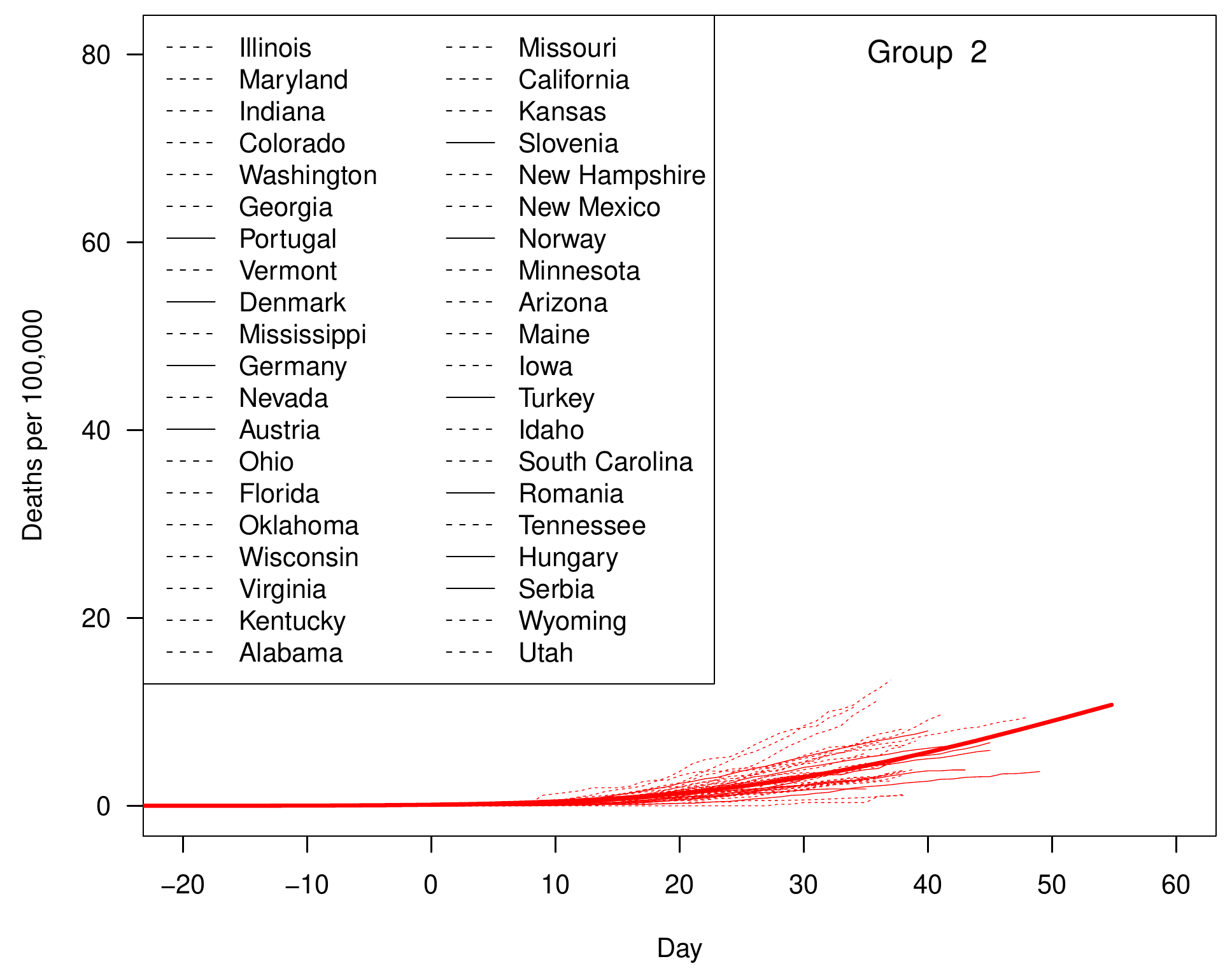}}
       }
   \mbox{
    	   	\subfigure{\includegraphics[angle=0,totalheight=2.3in]{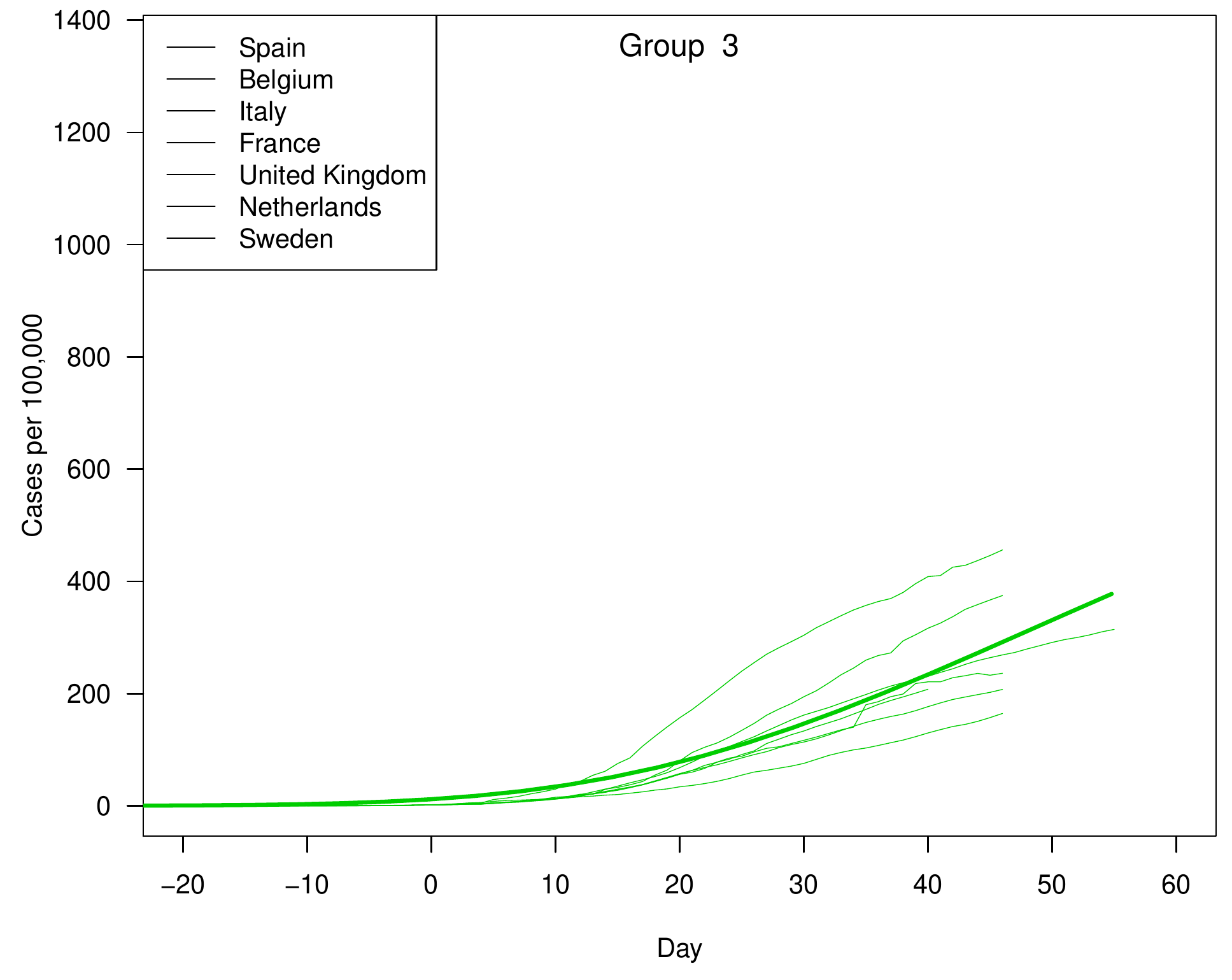}}
          	\subfigure{\includegraphics[angle=0,totalheight=2.3in]{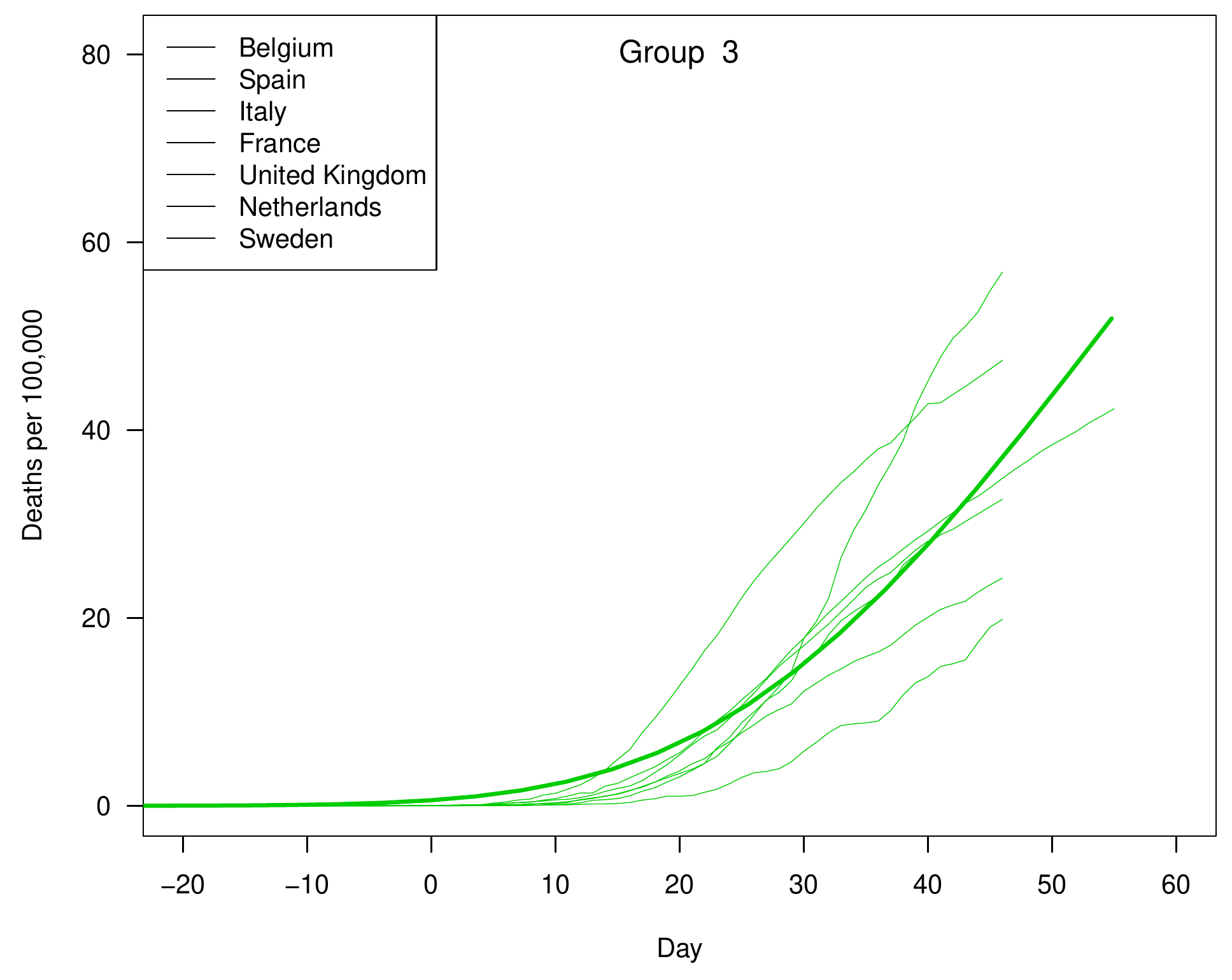}}
        }
 \caption{Individual cluster memberships for population adjusted cumulative cases and deaths (Groups 1 - 3).}
  \label{fig.groups1}
 \end{figure}
 
 \begin{figure}[t!]\ContinuedFloat
\centering
 \mbox{
 	\subfigure{\includegraphics[angle=0,totalheight=2.3in]{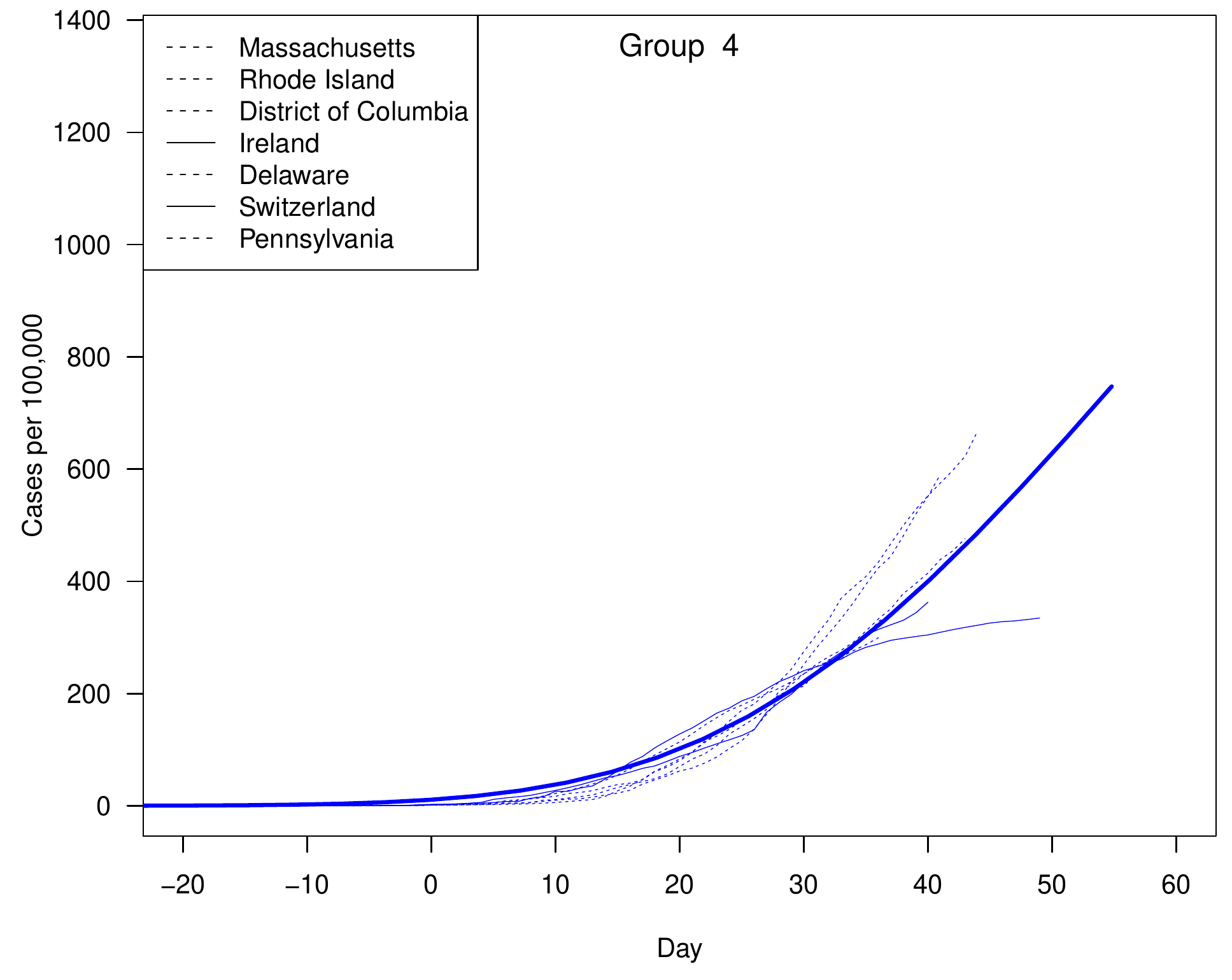}}
 	\subfigure{\includegraphics[angle=0,totalheight=2.3in]{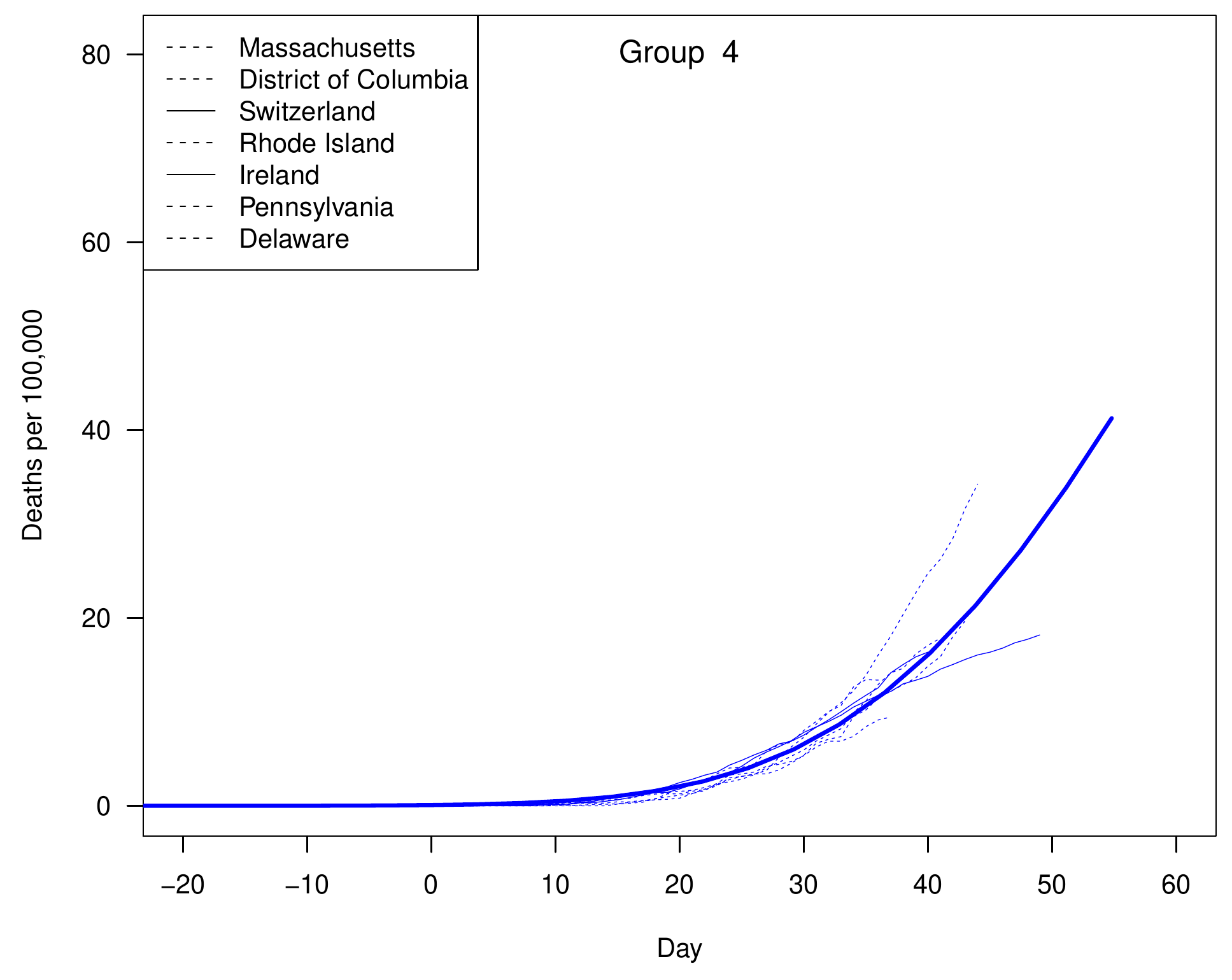}}
 }
 \mbox{
 	\subfigure{\includegraphics[angle=0,totalheight=2.3in]{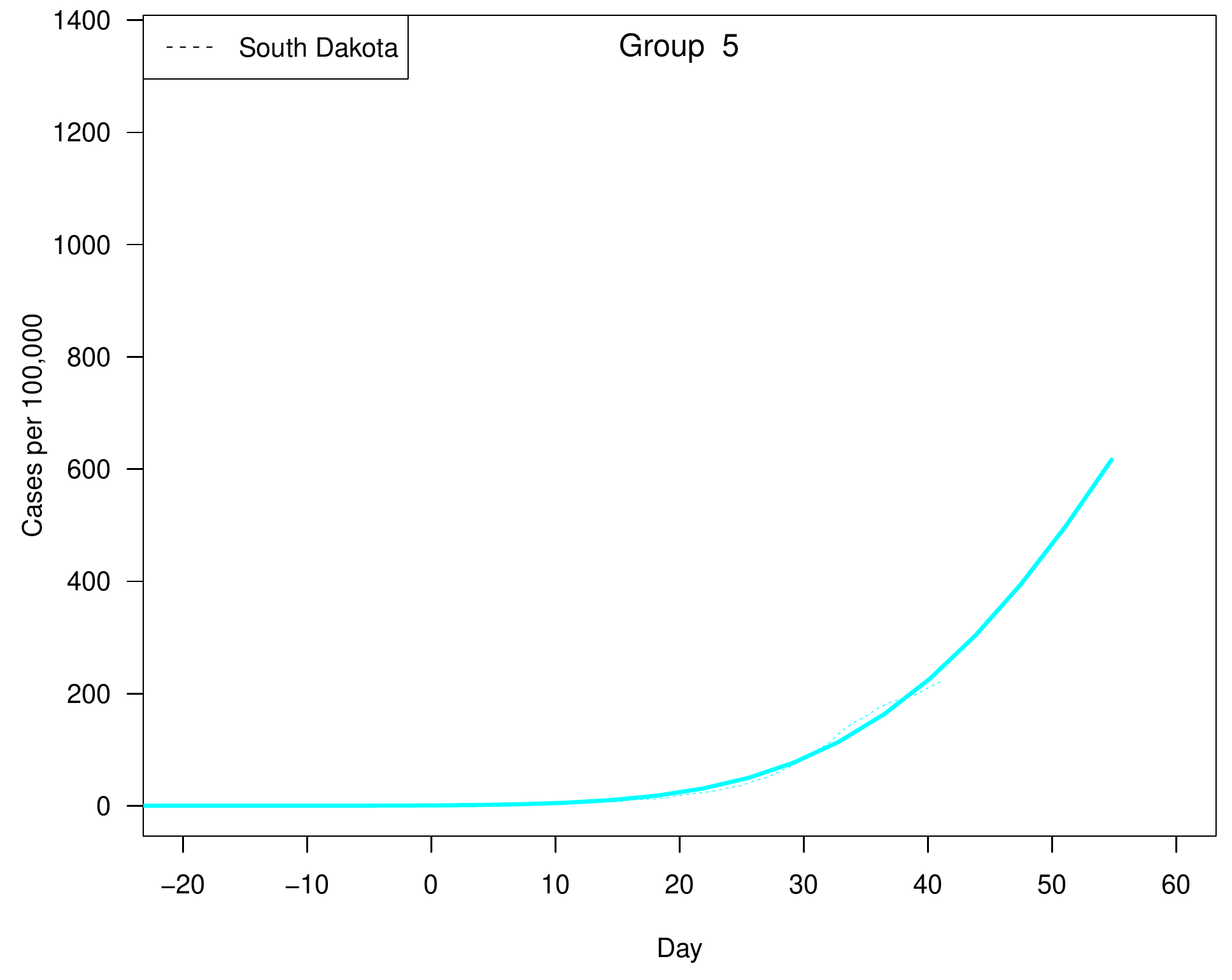}}
 	\subfigure{ \includegraphics[angle=0,totalheight=2.3in]{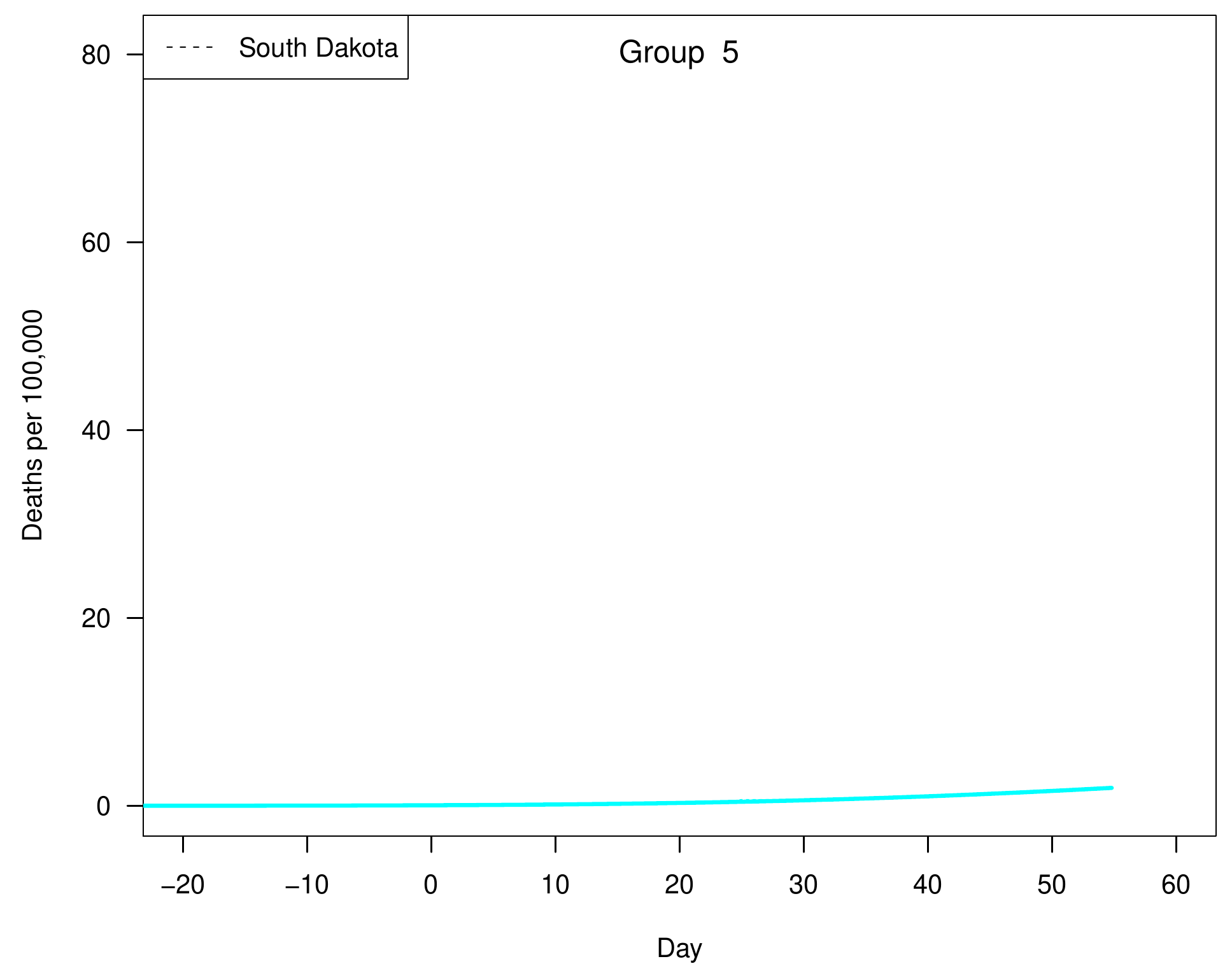}}
 }
 \mbox{
 	\subfigure{\includegraphics[angle=0,totalheight=2.3in]{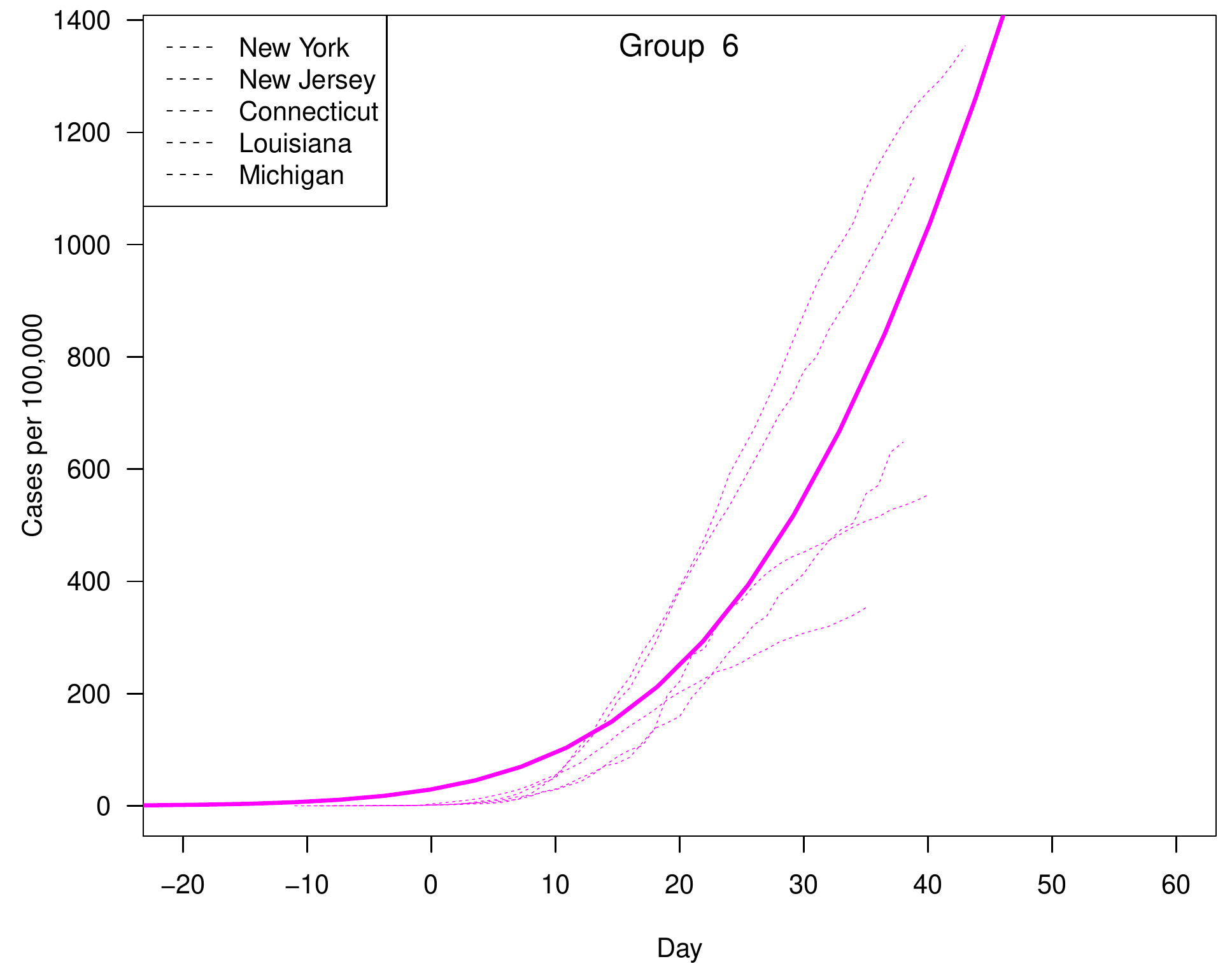}}
 	\subfigure{\includegraphics[angle=0,totalheight=2.3in]{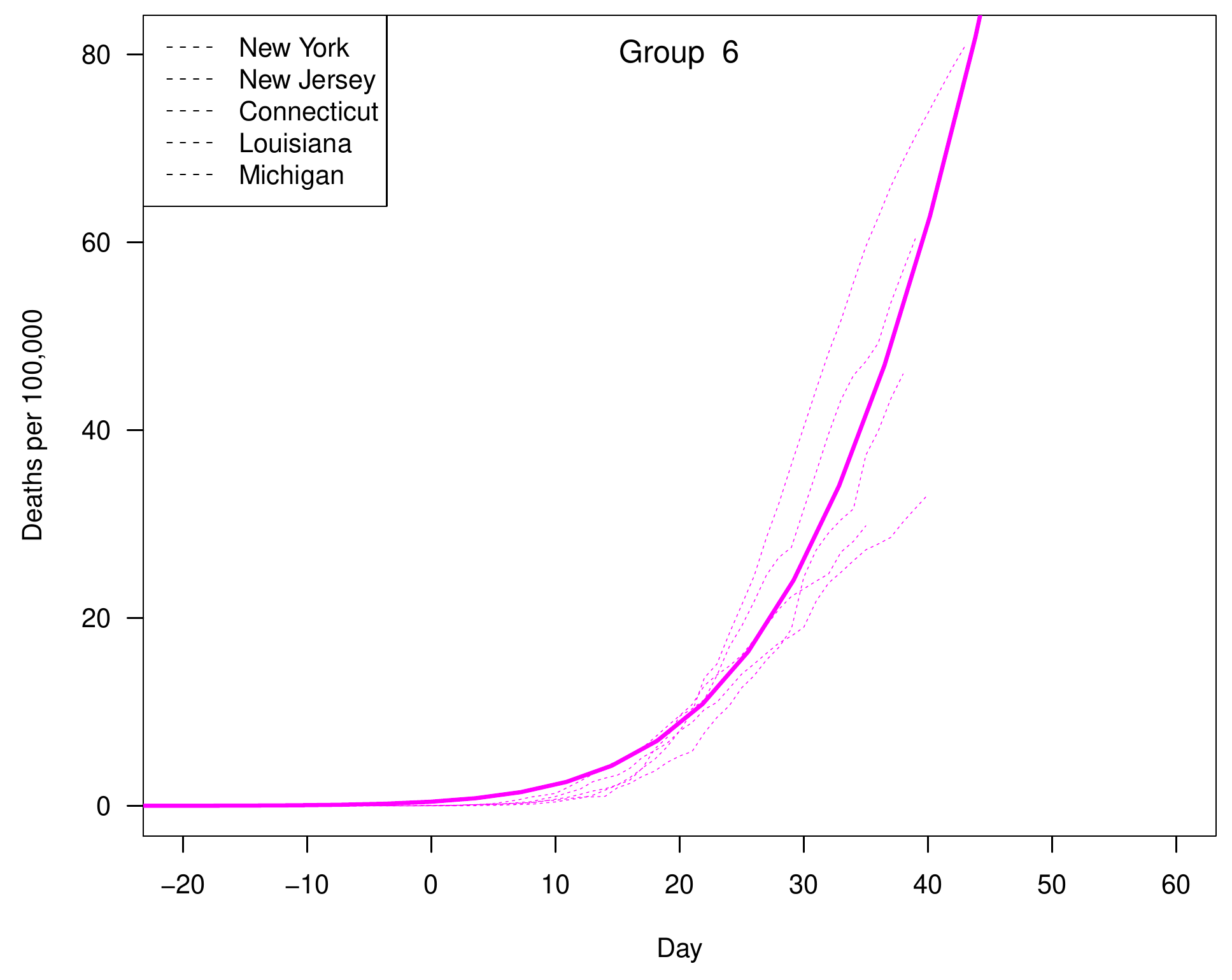}}
 }
   \caption[]{(cont.) Individual cluster memberships for population adjusted cumulative cases and deaths (Groups 4 - 6).}
  \label{fig.groups2}
 \end{figure}

\section{Conclusion}
\label{sect.disc}
A semi-supervised clustering model is used for identifying similar trends in the number of cases and deaths for 23 European countries and 51 US states. The mixture of non-linear regression models based on generalized logistic function proved to be flexible and identified regions with similar trends of infection and death rates. This model provides insights into how different regions reacted to this global crisis, which of them have successfully mitigated against its unmanageable spread, and which of them are in a worse state. The dataset contained most European countries that were exposed to the virus earlier than the majority of US states. As a result, we can learn how each group's trends differ and use this analysis to gain insight.

Although, members of the same group show similar trends, the regions exhibit broad variation in demography, geography, and social behaviour - factors that are known to influence the transmission rate \citep{cdcgeo20}. Their inclusion in a group appears to be due to a combination of mitigation strategies, time of onset, delay in mitigation, and demography. A positive case in point is the state of California with its large population, relatively higher density, earlier time of onset assigned in one of the groups with good trends. The effect of demography is made clear in the case of South Dakota where the death rate is very low owing to the majority of its cases being younger than the national average. On the other hand, the opposite effect of demography is also apparent in Group 3 where a higher mortality rate relative to their infection rate is observed. These countries are characterized by an aging population with 18 - 22\% greater than 65 years old \citep{worldbank20}. The trend in the most impacted group that is comprised of New York, New Jersey, and others may be a combination of delayed mitigation, high population density, and socio-cultural circumstances (as in the case of Louisiana's spiked cases due to an annual gathering 'Mardi Gras'\citep{cdcgeo20}). In addition, a recent report has identified that the virus arrived in New York weeks ahead of the first case from Europe. It appears that latent infection may have given the virus foothold by the time the first case was officially announced.

Overall, each cluster exhibits unique characteristics that can, in most cases, be explained by the composition of its members. One of the reasons for this grouping was joint modeling of the death and case rates. The nonlinear regression models within each cluster can be extrapolated to give a precise estimate of where a given region might be in a few days or weeks if they continue with the same strategy and as a result the same trajectory. In particular, we hope this analysis may help provide an insight to make a data-driven decision. Generally, the groupings can help regions to make strategic decisions on possible changes in mitigation and reopening or relaxing of restrictions. The model can be implemented to look into behaviors of other parts of the world to help develop public health policies by learning from data and experience.

The main limitations of this analysis are the reliability of the reported cases and deaths and inconsistencies in testing practices. Therefore, some of the trends and hence the groupings might be an artifact of the above. Additionally, to bring the regions to the same scale, the time of onset was chosen to be the date one region reaches 1 case per 100,000. This was a reasonable but somewhat arbitrary choice and results might change slightly depending on the choice of this date. Finally, the factors that may be related to high or low infection and death rates are not extensively explored. Further studies are needed to assess different factors including those that are inherent ({\it e.g.,} geographic location, population density, or demographic structure) and  mitigation strategies that can ameliorate these factors.

\section{Data Availability Statement}
The COVID-19 related data used in this paper are publicly available at  https://cran.r-project.org/web/packages/coronavirus/index.html and The New York times GitHub page https://github.com/nytimes/covid-19-data. The population data are available at https://www.census.gov/programs-surveys/popest/data/data-sets.html and https://data.worldbank.org/indicator/sp.pop.totl

%\nocite{*}% Show all bib entries - both cited and uncited; comment this line to view only cited bib entries;
%\bibliography{wileyNJD-APA}

\end{document}